\documentclass[usenatbib]{aa} 

\usepackage{graphicx}   
\usepackage{amsmath}    
\usepackage{amssymb}    
\usepackage{gensymb}
\usepackage{txfonts}
\usepackage[colorlinks=true]{hyperref}
\usepackage{soul}
\usepackage{array}
   
\defcitealias{Pierre:2016aa}{XXL~Paper~I}
\defcitealias{Pacaud:2016aa}{XXL~Paper~II}
\defcitealias{Adami:2018aa}{XXL~Paper~XX}
\defcitealias{Faccioli:2018aa}{XXL~Paper~XXIV}

\begin{document} 

  \title{The XXL Survey\\ 
  XLV. Linking the ages of optically selected groups to their X-ray emission}
  
  \titlerunning{GAMA/XXL: X-ray emission in optically selected groups}

  \author{
  J. P. Crossett$^{1,2}$\thanks{email: jacob.crossett@uv.cl}
  \and
  S. L. McGee\inst{2}
  \and
  T. J. Ponman\inst{2}
  \and
  M. E. Ramos-Ceja\inst{3}
  \and
  M. J. I. Brown\inst{4}
  \and
  B. J. Maughan\inst{5}
  \and
  A. S. G. Robotham\inst{6}
  \and  
  J. P. Willis\inst{7}
  \and
  C. Wood\inst{5}
  \and
  J. Bland-Hawthorn\inst{8}
  \and
  S. Brough\inst{9}
  \and
  S. P. Driver\inst{6}
  \and
  B. W. Holwerda\inst{10}
  \and
  A. M. Hopkins\inst{11}
  \and
  J. Loveday\inst{12}
  \and
  M. S. Owers\inst{13,14}
  \and
  S. Phillipps\inst{5}
  \and
  M. Pierre\inst{15}
  \and
  K. A. Pimbblet\inst{16}
  \\
     }
  \authorrunning{Crossett et al.}
  
  \institute{
  Instituto de F\'{i}sica y Astronom\'{i}a, Universidad de Valpara\'{i}so, Avda. Gran Breta\~{n}a 1111 Valpara\'{i}so, Chile
  \and
  School of Physics and Astronomy, University of Birmingham, Birmingham, B15 2TT, UK
  \and
  Max-Planck Institut f\"{u}r extraterrestrische Physik, Postfach 1312, 85741 Garching bei M\"{u}nchen, Germany
  \and
  School of Physics and Astronomy, Monash University, Clayton, VIC 3800, Australia
  \and
  H. H. Wills Physics Laboratory, University of Bristol, Tyndall Ave, Bristol BS8 1TL, UK
  \and
  ICRAR, University of Western Australia, Crawley, WA 6009, Australia
  \and
  Department of Physics and Astronomy, University of Victoria, 3800 Finnerty Road, Victoria, V8P 5C2, BC, Canada
  \and
  Sydney Institute for Astronomy, School of Physics, University of Sydney, NSW 2006, Australia
  \and
  School of Physics, University of New South Wales, NSW 2052, Australia
  \and
  Department of Physics and Astronomy, 102 Natural Science Building, University of Louisville, Louisville KY 40292, USA
  \and
  Australian Astronomical Optics, Macquarie University, 105 Delhi Rd, North Ryde 2113, Australia
  \and
  Astronomy Centre, University of Sussex, Falmer, Brighton, BN1 9QH, UK
  \and
  Department of Physics and Astronomy, Macquarie University, NSW, 2109, Australia
  \and
  Astronomy, Astrophysics and Astrophotonics Research Centre, Macquarie University, Sydney, NSW 2109, Australia
  \and
  AIM, CEA, CNRS, Université Paris-Saclay, Université Paris Diderot, Sorbonne Paris Cité, F-91191 Gif-sur-Yvette, France
  \and
  E.A. Milne Centre for Astrophysics, University of Hull, Cottingham Road, Kingston-upon-Hull, HU6 7RX, UK
  }

\date{Received XXX; accepted YYY}
\abstract{Why are some galaxy groups pervaded by a hot X-ray emitting intracluster medium, whilst others have no detectable X-ray emission? Is the presence of hot gas a reliable indicator of dynamical maturity, and can some virialised groups contain little or none of it? What are the main differences between samples of groups selected in the X-ray and optical bands? We address these questions by studying 232 optical spectroscopically selected groups from the Galaxy And Mass Assembly (GAMA) survey that overlap the XXL X-ray cluster survey. X-ray aperture flux measurements combined with GAMA group data provides the largest available sample of optical groups with detailed galaxy membership information and consistently measured X-ray fluxes and upper limits.

A sample of 142 of these groups is divided into three subsets based on the relative strength of X-ray and optical emission, and we see a trend in galaxy properties between these subsets: X-ray overluminous groups contain a lower fraction of both blue and star forming galaxies compared with X-ray underluminous systems.
X-ray overluminous groups also have a more dominant central galaxy, with a magnitude gap between first and second ranked galaxies on average 0.22 mag larger than in underluminous groups. Moreover, the central galaxy in overluminous groups lies closer to the luminosity-weighted centre of the group. We examine a number of other structural properties of our groups, such as axis ratio, velocity dispersion, and group crossing time, and find evidence of trends with X-ray emission in some of these properties despite the high stochastic noise arising from the limited number of group galaxies.

We attribute the trends we see primarily to the evolutionary state of groups, with X-ray overluminous systems being more dynamically evolved than underluminous groups. The X-ray overluminous groups have had more time to develop a luminous intragroup medium, quench member galaxies, and build the mass of the central galaxy through mergers compared to underluminous groups.

However, an interesting minority of X-ray underluminous groups have properties that suggest them to be dynamically mature. We find that the lack of hot gas in these systems cannot be accounted for by high star formation efficiency, suggesting that high gas entropy resulting from feedback is the likely cause of their weak X-ray emission.
}


\maketitle
%

\section{Introduction}
\label{Section:Intro}
In the low redshift universe, small collections of galaxies, such as galaxy groups, are the modal environment in which galaxies reside \citep[e.g.][]{Eke:2005aa, Aragon-Calvo:2010aa}. Group environments provide excellent laboratories in which to study the processes of environmentally driven galaxy transformation, since galaxies can experience interactions with other member galaxies, as well as with the group potential and the intergalactic gas it holds. Investigating how galaxies evolve within groups as these group structures themselves form and develop is vital to our understanding of galaxy evolution. 

It is well established that environment is an important driver of galaxy evolution \citep[e.g.][]{Oemler:1974aa, Dressler:1980aa, Peng:2010aa, Peng:2012aa}. High-density environments are linked with changes in a number of galaxy properties, including an increase in the galaxy red fraction \citep[e.g.][]{Balogh:2004aa, Blanton:2005aa, Weinmann:2006aa, Baldry:2006aa, van-den-Bosch:2008aa, Bamford:2009aa} and a lower average galaxy star formation rate \citep[e.g.][]{Lewis:2002aa, Gomez:2003aa, Kauffmann:2004aa, von-der-Linden:2010aa}. The long-established morphology-density relation also links environment to the morphological mix of galaxies, with the modal galaxy morphology shifting from spiral to elliptical in regions of high galaxy density \citep[e.g.][]{Oemler:1974aa, Davis:1976aa, Dressler:1980aa, Postman:1984aa, Dressler:1997aa, Bamford:2009aa, van-der-Wel:2010aa}.

There are numerous ways in which environments such as groups and clusters can cause such transformations in galaxy populations. Group-sized haloes, with low relative velocities between member galaxies, can facilitate mergers and tidal interactions, disrupting the structure of galaxies and creating tidal features \citep[e.g.][]{Ghigna:1998aa, Moore:1999aa, Hernandez-Fernandez:2012aa}. At the higher relative velocities present in rich clusters, rapid galaxy fly-by encounters can perturb galaxy structure and change galaxy properties in a process known as {`harassment}' \citep[e.g.][]{Moore:1996aa, Porter:2008aa, Aguerri:2009aa}.

In addition to interactions between member galaxies, groups and clusters also contain a hot intergalactic plasma known as the intracluster medium or the intragroup medium (IGM). This hot IGM interacts with the gas present in infalling galaxies, causing the removal of the hot gaseous halo surrounding the galaxy \citep[sometimes called strangulation or starvation;][]{Larson:1980aa, Balogh:2000aa} or, in more severe cases, the removal of the cold gas disk \citep[known as ram-pressure stripping; e.g.][]{Gunn:1972aa, Abadi:1999aa, Chung:2007aa, Owers:2012aa, Poggianti:2016aa, Barsanti:2018aa, Schaefer:2019aa}. A combination of these processes is thought to lead to the high fraction of quiescent, spheroidal galaxies found within large galaxy groups and clusters. 

A variety of different techniques have been used to find and define galaxy groups and clusters. The nature of the group sample obtained, and hence the environmental influences on the member galaxies, can vary according to the method used. This has been responsible for a good deal of confusion and disagreement between different studies of galaxy-environment relationships. For example, selecting groups via X-ray emission may privilege virialised structures, in which the IGM has been heated by collapse into a well-developed group potential, and may miss smaller, younger groups that could be found using an optical selection method \citep[e.g.][]{Finoguenov:2009aa, Dietrich:2009aa}. Optical selection of groups can employ a variety of techniques: finding visual overdensities \citep[e.g.][]{Abell:1957aa, Dalton:1997aa}, through red sequence fitting \citep[e.g.][]{Gladders:2000aa, Gladders:2005aa, Andreon:2011aa}, or by analysing spectroscopic galaxy redshift surveys using a clustering or friends-of-friends (FoF) algorithm \citep[e.g.][]{Eke:2004aa, Brough:2006aa, Robotham:2011aa, Tempel:2014aa}. However, these optically based methods do not give a measure of the state of the IGM, which requires X-ray observations, and optical surveys can be vulnerable to contamination by apparent groups that are actually line-of-sight superpositions of galaxies \citep[e.g.][]{McNamara:2001aa, Barkhouse:2006aa} rather than truly bound systems. Obtaining a statistically homogeneous sample that combines optical groups with consistently measured X-ray luminosities has so far proven to be an elusive goal.

Previous studies that have measured the X-ray content of optically selected groups have found these systems to be X-ray underluminous for their mass compared to X-ray selected systems (\citealt{Bower:1994aa, Castander:1994aa, Gilbank:2004aa, Lubin:2004aa, Fang:2007aa, Dietrich:2009aa, Andreon:2019aa}; see also \citealt{Andreon:2016aa}) and are sometimes undetected in X-rays, especially in clusters at high redshift or low richness \citep[e.g.][]{Sadibekova:2014aa, Rozo:2014aa}. This difference in the X-ray emission between optical and X-ray selected samples indicates that groups of a given optical richness can have a significant range in X-ray luminosity, pointing to systematic differences in the properties of the hot IGM. 

Several theories have been put forward to explain the origin of such variations between groups in the IGM. Firstly, a dynamically young group or cluster will still be forming and will therefore be incompletely virialised. In such a situation, the intracluster medium has not yet been fully compressed in the developing potential well to form the hot, high-density plasma that generates X-ray emission \citep[e.g.][]{Donahue:2001aa, Gilbank:2004aa, Barkhouse:2006aa, Rasmussen:2006aa, Popesso:2007aa, Balogh:2011aa}. 

Alternatively, energy injection into the IGM from supernovae or active galactic nuclei (`feedback') may have raised the entropy of the IGM above the value generated by shocks during group collapse. An increased IGM entropy results in a lower central density of the IGM once it comes into hydrostatic equilibrium in the potential well, thus lowering the X-ray emission \citep[e.g.][]{Bower:1997ab, Gilbank:2004aa, Hicks:2008aa, McCarthy:2010aa, Pearson:2017aa}. In this picture, the X-ray luminosity of a group will be influenced by its feedback history.

Finally, it is possible that galaxies within some groups could have very high star formation efficiency. In such a case, more of the gas would be converted into stars, leaving less available for the IGM. This would lower the IGM density and hence the X-ray luminosity \citep[e.g.][]{Gilbank:2004aa, Hicks:2008aa}. Establishing the relative importance of these three possible processes is key to understanding the differences between group samples selected in different ways.

Previous studies investigating the range in X-ray properties of optically selected groups have been hampered by incomplete knowledge of group membership, poor characterisation of X-ray dim groups, and small samples of groups with both X-ray and optical data \citep[e.g.][]{Finoguenov:2009aa, Andreon:2011aa}. Many previous studies that targeted optically selected groups used only shallow ROSAT X-ray measurements, with many groups and clusters simply undetected in X-rays \citep[e.g.][]{Donahue:2001aa, Mulchaey:2003aa, Brough:2006aa, Popesso:2007aa, Wang:2014aa, Roberts:2016aa}. Others had deeper X-ray measurements, but for only a small number of systems \citep[e.g.][]{Rasmussen:2006aa, Andreon:2011aa, Hicks:2013aa, Pearson:2017aa}. Furthermore, many comparisons of the optical and X-ray properties of groups have lacked the extensive optical spectroscopy needed to obtain redshifts for a statistically useful sample, with studies involving $<20$ groups in many cases \citep[e.g.][]{Gilbank:2004aa, Balogh:2011aa}. 

In order to make statistical comparisons between groups with different IGM properties, one requires a deep, spectroscopic, optical group catalogue with overlapping X-ray photometry across the entire field. It is only now, combining a large spectroscopically complete sample from the Galaxy and Mass Assembly (GAMA) survey \citep[][]{Driver:2011aa, Liske:2015aa, Baldry:2018aa} with well-matched X-ray data from the \textit{XMM-Newton} XXL survey \citep[XXL, ][hereafter \citetalias{Pierre:2016aa}]{Pierre:2016aa}, that such a study is possible.

Using this combined group sample, we investigate the properties of optically selected galaxy groups separated into different subsets based on their X-ray luminosity. In this way we aim to explore the relative contribution of different factors that may affect the state of the IGM in galaxy groups. If a galaxy group is X-ray underluminous due to dynamical youth, then we should see evidence of this in the structure of the group or the properties of its galaxies \citep[e.g.][]{Ribeiro:2013aa, Roberts:2018aa, Yuan:2020aa}. Alternatively, a high stellar mass fraction in X-ray underluminous groups would point to variations in star formation efficiency. If neither of these factors seems dominant, then variations in the impact of feedback from group to group could be indicated.

The structure of the paper is as follows: In Sect. \ref{Section:Data} we describe the sources of optical and X-ray data, including the method for deriving X-ray luminosities and upper limits. Section \ref{Section:sample} details the process of separating our sample into three different sub-samples based on X-ray luminosity. In Sect. \ref{Section:Results} we present our analysis, comparing the properties of these group sub-samples. We discuss our findings and the evolution of X-ray under- and overluminous groups in Sect. \ref{Section:Discussion} and draw our conclusions in Sect. \ref{Section:Summary}. Throughout this work we use AB magnitudes and adopt a cosmology consistent with prior XXL survey studies, based on measurements from the Wilkinson Microwave Anisotropy Probe satellite mission \citep{Hinshaw:2013aa}, with values $\Omega_\textrm{m}=0.28$, $\Omega_{\Lambda}=0.72$, and $H_{0}=70 ~\textrm{km}~\textrm{s}^{-1}~\textrm{Mpc}^{-1}$. Our optical luminosities are all expressed in units of solar luminosity.

\section{Data}
\label{Section:Data}
\subsection{Galaxy And Mass Assembly}
In this work we use data from the GAMA survey \citep[][]{Driver:2011aa, Liske:2015aa, Baldry:2018aa}, which provides optical spectra for $\sim$300,000 galaxies over five different regions in the sky. The multi-pass optical spectroscopy is $98\%$ complete to a magnitude of $r =19.8$ mag over three primary equatorial regions of the sky, even within dense concentrations of galaxies, making it ideal for probing environments such as groups \citep[][]{Robotham:2010aa, Gordon:2018aa}. 
One additional field in the GAMA survey, designated G02, has slightly lower completeness but has a significant overlap with the north field of the XXL survey \citepalias[][see our Sect. \ref{Section:XXL}]{Pierre:2016aa}, giving high quality X-ray photometry. GAMA provides redshifts for $\sim$35,000 galaxies in the $\sim$56 deg$^2$ of the G02 field, between right ascensions $30.2^{\degree}$ to $38.8^{\degree}$ and declination $-10.25^{\degree}$ to $-3.72^{\degree}$. The GAMA survey includes additional auxiliary information, including spectral line strengths and group memberships for all galaxies in this region \citep[][]{Robotham:2011aa, Gordon:2017aa}. More information about the G02 field is be found in \citet{Baldry:2018aa}. GAMA spectroscopy in this field is over 95$\%$ complete for 19.6 deg$^2$ that overlaps the XXL north field, between right ascension $30.2^{\degree}$ to $38.8^{\degree}$ and declination $-6^{\degree}$ to $-3.72^{\degree}$. Within this region GAMA provides high quality redshifts \citep[defined as a redshift quality nQ $\geq 3$; see][for details]{Liske:2015aa}, for $\sim$20,000 galaxies, allowing galaxy groups to be identified as described in the next section.

\subsubsection{Group membership}
The extensive redshift survey has been used to construct a catalogue of galaxy groups within the three equatorial fields as well as the G02 field in the GAMA survey \citep{Robotham:2011aa, Baldry:2018aa} using a FoF algorithm. This algorithm identifies galaxies that are neighbours in both sky position and velocity space, and links them to construct groups of galaxies with at least two members. The catalogue includes basic group parameters, such as group radius, axial ratio, and velocity dispersion, for all groups with at least five members. Additionally, properties such as the group dynamical mass are calculated, and total $r$-band optical luminosities are derived for each group, including a correction for flux missed due to the magnitude limit of the survey. An extensive set of mock catalogues, based on cosmological simulations, has been used to calibrate the grouping algorithm and assess any biases in derived group properties. Further details can be found in \citet{Robotham:2011aa}.

The catalogue defines a centroid position for each group, using an iterative centre of light method. This proceeds by calculating the centre of light location using all group galaxies, then rejecting the member galaxy furthest from this position. This process is then repeated until only two members remain, whereupon the brightest of the two remaining members is taken as the centre of the group. This procedure was found in \citet{Robotham:2011aa} to best recover the group centre in mock catalogues. In some cases, however, this position may not coincide with the X-ray centre of the group. We discuss this point in Sect. \ref{section:centre}.

In this study, we make use of the most recent version of the group catalogue, G$^{3}$Cv10, and restrict our sample to groups with at least five member galaxies above the magnitude limit. This catalogue contains 2540 groups within the G02 region, of which 307 contain at least 5 members above the GAMA apparent magnitude limit of $r =19.8$ mag. Of these groups, 238 lie within the region overlapping the XXL footprint (Fig. \ref{fig:XXLfields}; see Sect. \ref{Section:XXL} for more details). These groups have a mean of $\sim$9 group members, and a mean optical ($r$-band) luminosity of $\sim 1.5 \times 10^{11}$ L$_{\odot}/h^{2}$. Our study is based on these GAMA groups, taking the iterative centre of each group as the reference point for extracting X-ray photometry. 
\label{Section:GAMA}

\subsection{XXL}
\label{Section:XXL}
The XXL survey is one of the largest {\it XMM-Newton} surveys, totalling nearly 7 Ms of XMM observations covering 50 deg$^2$ with $\sim600$ pointings covering two large areas of the sky. This survey provides a sensitivity of $6\times 10^{-15}$ erg s$^{-1}$ cm$^{-2}$ in the [$0.5-2$]~keV energy band for point sources. The north field (XXL-N) covers an area largely contained within right ascension $30^{\degree}$ to $39^{\degree}$ and Declination $-6\degree$ to $-4\degree$. This provides a significant overlap with the GAMA G02 region, with the overlapping region containing 238 groups for use in this study.

\begin{figure*}
\begin{center}
\includegraphics[width=\textwidth]{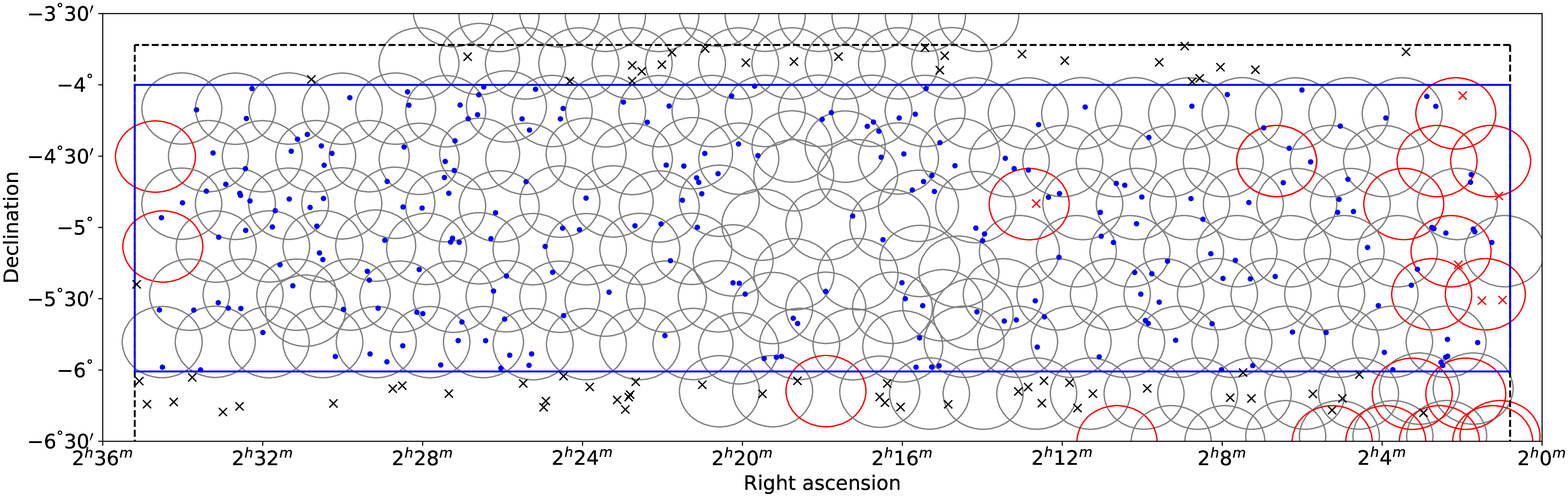}
\caption{Locations of the FoF group centres (blue points) with respect to the XXL fields (grey and red circles). The GAMA field used in this study is shown with a blue box, taken as a subset from the G02 field (dashed black lines). We remove six optical groups (red crosses) that lie in fields with high sky backgrounds (red circles), in addition to groups outside our matching area (black crosses). The remaining 232 groups (blue points) are taken for this analysis (see Table \ref{table:Xray_samples}  for details).}
\label{fig:XXLfields}
\end{center}
\end{figure*}

When combining these data, we ensured that all groups in our sample have high quality X-ray data and hence removed six groups that lie within fields with high XMM background counts. Figure \ref{fig:XXLfields} shows the configuration of the overlapping XXL pointings and highlights those with high sky backgrounds. This leaves 232 GAMA groups with at least five members that have good X-ray data from the XXL north field.

\subsubsection{X-ray aperture photometry}
\label{Section:XrayPhotometry}

While the XXL survey team has published several catalogues of extended X-ray sources (e.g. \citealt{Pacaud:2016aa}, hereafter \citetalias{Pacaud:2016aa}, \citealt{Adami:2018aa}, hereafter \citetalias{Adami:2018aa}), these catalogues only record the brighter X-ray sources. For our study, we required X-ray measurements or upper limits for {all} optical groups in our sample, and we are especially interested in those that have low X-ray luminosity.
In order to achieve this we extract our own X-ray photometry from a region of radius 300 kpc about each GAMA group centre using the method described in \citet{Willis:2018aa}. Briefly, we compute the background-marginalised posterior probability distribution function (pdf) of the source count-rate. For this, photon counts and exposure time information of the source and background aperture are extracted. The photon counts are described by Poisson likelihoods. 
If the source lies $<2'$ from the pointing centre, the background aperture is a detached annulus ($1'$ outside the source aperture) of width $1.5'$ centred on the group position. When the group lies $>2'$ from the pointing centre, the background is taken from an annulus centred on the pointing direction, which has a width equal to the diameter of the source circle, and a radius equal to the off-axis angle of the source. This annulus encompasses the source aperture, but a region around the source is excluded from it. All point-like sources detected by the XXL pipeline are masked out from each image, exposure, and background map along with any detector gaps and bad detector regions (see \citetalias{Pacaud:2016aa} for details). 

The method is applied individually to each of the three {\it XMM-Newton} cameras, giving three count-rate posterior probability distributions per source in the $0.5-2$~keV energy band. Part of the 300 kpc source aperture may be lost for four reasons: (i) the source circle may extend beyond the sensitive area of the detector array, (ii) all three EPIC cameras have gaps between charge-coupled device (CCD) chips in the focal plane, (iii) some regions may be masked where contaminating point sources have been removed, and (iv) one of the CCD chips in the MOS1 camera ceased to operate part way through the mission and has effectively been dead since 2005. We applied a simple geometric scaling to account for any area lost from the source aperture due the combination of these effects. However, where any of the three {\it XMM-Newton} cameras had more than $50\%$ of the aperture lost, the flux measurement from that camera has been removed from the calculation of the final posterior distribution.

In two cases, photometry within the 300 kpc aperture was not possible due to artefacts and bright point sources close to the source position. A further 16 groups had masking covering more than $50\%$ of the source aperture in all three {\it XMM-Newton} cameras. An additional single group was found to have highly discrepant flux values between the three {\it XMM-Newton} detectors. All of these systems were removed from the sample, leaving 213 groups in our group sample. 

Count rates for each camera are converted into a k-corrected source-frame flux using an energy conversion factor, calculated with X-Ray Spectral Fitting Package XSPEC \citep{Arnaud:1996aa} using an APEC emission model with temperature $T=1$~keV, column density $N_\textrm{H}=2.6\times 10^{20}$ cm$^2$, metallicity $Ab = 0.3$, and standard on-axis EPIC response matrices. The group redshift of the GAMA optical counterpart is used for each source. This gives a flux pdf for each camera, and the final flux posterior pdf is obtained by multiplying the available individual distributions.
The resultant posterior distribution is then used to determine the X-ray flux of each group. The peak of the pdf gives a best flux estimate for all sources where a peak is found, and the pdf is also used to calculate an asymmetric one sigma error bar.
In some cases the pdf is highest at zero flux. In this case we instead calculate a $90\%$ flux upper limit from the posterior distribution. The flux measurements and upper limits are used, in conjunction with the GAMA-derived redshift, to calculate an X-ray luminosity for each group, denoted  L$^{XXL}_{300kpc}$.

\subsubsection{Aperture size effects}
We employed a metric aperture for this study, rather than extracting X-ray flux within an overdensity radius, due to its simplicity (allowing easy comparison with later studies) and independence of the assumptions required to estimate system mass. Compared to a metric radius, use of an overdensity radius would systematically raise the L$_X$ values of more massive systems relative to less massive ones, resulting in a somewhat steeper relationship between L$_X$ and L$_{opt}$ than that found below. However, the classification of groups into X-ray overluminosity classes (see Sect. \ref{Section:sample}), which is our main aim in this study, should be essentially unchanged.

The choice of a 300 kpc radius aperture is intended to encompass the core region of all the groups in our sample. If the aperture is too large compared to the source it will contain a higher fraction of background flux, and consequently have higher measurement uncertainties. Conversely, a smaller aperture may miss significant flux from the extended hot gas.

We explored the use of two aperture sizes: 100 kpc and 300 kpc. We found that the two flux measurements generally scale with each other, albeit with some scatter. However, the smaller aperture unsurprisingly lost more source flux and also resulted in 16 sources with no reliable flux estimate due to high fractions of the aperture being lost to point source masking. The number of systems lost decreased to two when the larger 300 kpc aperture was employed. Given this, we proceed using flux values from the 300 kpc aperture, unless otherwise noted. We have, however, confirmed that using 100 kpc fluxes does not significantly change the results of our study. 

Since the correction for the lost area within the source aperture is a simple one, we conducted checks to explore whether point source masking might have adversely affected our L$^{XXL}_{300kpc}$ estimates. 
We find no correlation between L$^{XXL}_{300kpc}$ and the fraction of the aperture masked, or the number of masked point sources. This demonstrates that the masking of point sources within our X-ray apertures is not a major cause of the scatter in L$^{XXL}_{300kpc}$ (we recall that groups with over $50\%$ masking have already been removed from our sample). However, it is still possible that masking could have an impact in some individual cases where point sources or chip gaps lead to significant masking in the inner regions of the source aperture. Such cases are discussed in Sect.\ \ref{Section:biases}.

\subsection{Comparison with existing X-ray clusters}
\label{Section:CompareXXL}
X-ray luminosities and positions for the brightest 365 extended sources in the XXL survey have already been published in \citetalias{Adami:2018aa}.
In the following, we compare our results with the 74 sources from this cluster catalogue that lie within the field and redshift range of our sample. This is done for three reasons: the comparison provides a check on both our calculated luminosities and group centre positions, and it allows us to check for groups that may be part of larger source complexes. 

\subsubsection{Removal of complex systems}
Since our main purpose is to study the properties of {individual} galaxy groups, we need to remove groups that are part of larger cluster or filament systems. Such systems, including groups in the process of merging, or falling into a cluster, may be confused in terms of both galaxy membership and X-ray emission. The additional processes at work in dynamic large-scale structures can modify the properties of  member galaxies \citep[e.g.][]{Markevitch:1999aa, Poggianti:2004aa, Owers:2012aa, Kleiner:2014aa, Stroe:2020aa}, and the state of the IGM \citep[e.g.][]{Zabludoff:1995aa, Brough:2006ab, Bekki:2010aa, Owers:2011aa}. While these systems are, of course, of great interest in their own right, they represent an unwanted complication for the present study.

\begin{figure*}
\begin{center}
\includegraphics[width=\textwidth]{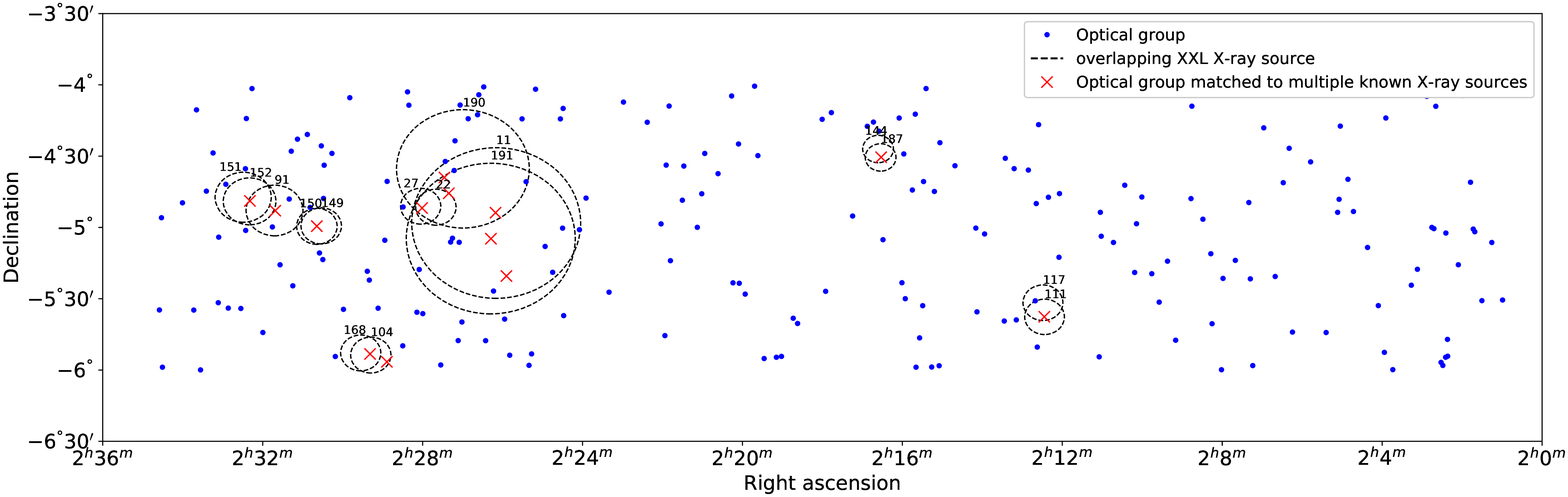}
\caption{Positions of the GAMA FoF groups compared to X-ray cluster complexes. X-ray clusters from \citetalias{Adami:2018aa} that overlap one another within 5000 kms$^{-1}$ and 2 Mpc are shown as black dashed circles with a radius of 2 Mpc at the X-ray source redshift. Red crosses show the positions of GAMA groups that match to these X-ray sources in redshift space. These systems may be infalling or merging into cluster complexes, which may affect their properties. They have therefore been excluded from our analysis.}
\label{fig:overlapgroups}
\end{center}
\end{figure*}

In order to identify which groups could be considered a part of a larger complex system, we compare the locations of our optical FoF groups to X-ray sources from the X-ray cluster catalogue of \citetalias{Adami:2018aa}. Drawing a redshift space volume of 2 Mpc and 5000~km~s$^{-1}$ around each of the catalogued X-ray clusters, we define a potential complex system to be any region where at least two of these volumes overlap. The value of $\pm$ 2Mpc was chosen to ensure that it is larger than the radius to the most distant FoF member in any group in our sample. The 5000~km~s$^{-1}$ offset is deliberately chosen to be large because the X-ray source redshifts are often based on a small number of redshifts and are thus uncertain. These wide regions are chosen to be conservative.
We mark these overlapping clusters in Fig. \ref{fig:overlapgroups} with black dashed circles. The radius of each circle is 2 Mpc at the X-ray source redshift. The XLSSC source number is given above each source. FoF optical groups are shown in blue, and those that match to any of the complex X-ray sources within 2 Mpc and 5000 kms$^{-1}$ are marked with red crosses. These optical groups are deemed to be part of these larger complex systems and have been removed from our sample. A total of 13 groups are removed in this way, leaving 200 in our sample (see Table \ref{table:Xray_samples} for further details). 

We have checked that including these groups does not significantly alter the results presented. However, by excluding these systems, we are more confident that merging cluster systems, and infalling groups are not a contaminant in our results. 

\subsubsection{Optical versus X-ray centre}
\label{section:centre}
\begin{figure}
\begin{center}
\includegraphics[width=0.5\textwidth]{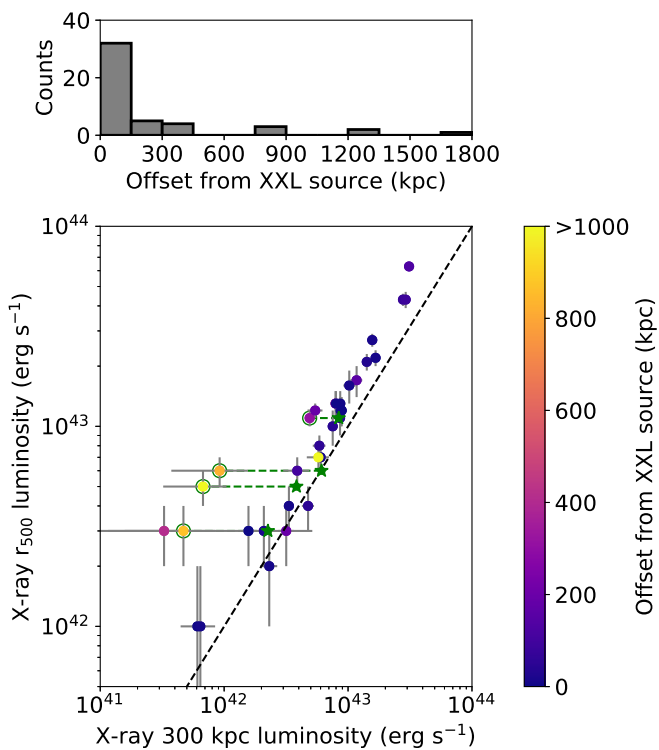}
\caption{XXL-derived [$0.5-2$]~keV L$^{XXL}_{r500,MT}$ from \citetalias{Adami:2018aa} plotted against our aperture [$0.5-2$]~keV L$^{XXL}_{300kpc}$ for the matched groups in our sample, with $1\sigma$ uncertainties shown. The colour bar displays the 2D projected sky offset between the GAMA group centre and the XXL X-ray position. The dotted black line denotes a 1:1 match. The two X-ray luminosity measurements are similar in most cases, except when there is a large offset between the XXL source position and the optical group centre. Green stars show re-measured L$^{XXL}_{300kpc}$ values for groups where a catalogue XXL source overlaps with a prominent non-central GAMA group member (see Sect. \ref{section:centre} for details). They are connected to the original values by a green line.}
\label{fig:XXLscalediff}
\end{center}
\end{figure}

In addition to removing complex systems, we also compare the flux measured from the XXL catalogue of \citetalias{Adami:2018aa} to our aperture X-ray photometry, to check for consistency between the two measurements. We find 47 optical FoF groups that match to a single source in the XXL catalogue within 3000 kms$^{-1}$ and 2 Mpc. We note that the 3000 kms$^{-1}$ is a smaller matching distance than we used in the previous section, but the optical group redshifts are based on more members and thus have smaller uncertainties. Source luminosities in the XXL catalogue have been calculated within an overdensity radius r$_{500}$, estimated from the X-ray temperature. For most groups this is larger than 300 kpc. In Fig. \ref{fig:XXLscalediff} our L$^{XXL}_{300kpc}$ aperture-derived X-ray luminosities are plotted against the Paper XX-derived X-ray luminosities (L$^{XXL}_{r500,MT}$) calculated within r$_{500}$. Both of these measurements are taken in the [$0.5-2$]~keV band. 

As can be seen, there is a good correlation between the two X-ray luminosities. Most points lie above the dashed line of equality, as expected given that r$_{500}$ is generally larger than $300$ kpc. However, a few sources deviate markedly, with a value of L$^{XXL}_{300kpc}$, which is substantially less than the corresponding L$^{XXL}_{r500,MT}$ value. The main reason for this can be deduced by examining the offset between the catalogue XXL source and the centre location of the matched optical GAMA group. This is shown by the colour of the points in Fig. \ref{fig:XXLscalediff}. It can be seen that all the most deviant points have a large offset between the GAMA group centre and the XXL source.

There are two possibilities for large spatial offsets of this sort. Either the X-ray source is associated with the GAMA group, but the X-ray emission is not centred on the galaxy located by the GAMA iterative centre of light algorithm, or the matched XXL source is not actually associated with the GAMA group at all. In the first case, we should move the centre of our X-ray flux aperture to best measure the emission, but in the second case we should not.

To distinguish between these two possibilities we studied the groups with large offsets individually. Nine GAMA groups with z$<0.35$ were found to match to a single XXL source with a projected separation of $>300$ kpc between the optical and X-ray centres. We visually inspected each of these groups to determine whether the XXL source is centred on a group member, or is located away from any group member. 

In four cases, the XXL source is not centred on any of the brighter GAMA group members, but appears to be associated with other galaxies in the foreground or background. In these cases, we assume the XXL source is not associated with the group, and keep the aperture flux centred on the GAMA iterative centre. We assume here that, given the distance between the XXL source and the GAMA source is $>300$ kpc, there is negligible contamination into our 300 kpc aperture.

In the five remaining groups, we find that the XXL source is indeed centred on a prominent member of the group.  We therefore re-calculated the aperture photometry for these five groups, relocating the centre of our aperture to the XXL source position. The aperture based  X-ray luminosities for these five systems rise by factors of between 1.7 and 6.6 when this adjustment is made, bringing them into much better agreement with the XXL catalogue luminosities. Four of these groups are highlighted in Fig.\ \ref{fig:XXLscalediff}, showing the original measured flux connected to the re-measured L$^{XXL}_{300kpc}$ value (green stars). The fifth group is not shown due to the source having poor L$^{XXL}_{r500,MT}$ signal-to-noise.
We use the adjusted L$^{XXL}_{300kpc}$ value for these five groups. In all other groups, the aperture X-ray photometry is calculated from the optical iterative group centre presented in the GAMA survey \citep{Robotham:2011aa}.

\section{X-ray overluminous and underluminous group selection}
\label{Section:sample}

\subsection{Metric for selection}
In order to examine the differences between groups with high X-ray emission and those with low X-ray emission, we need to define a metric for groups that are `X-ray overluminous' and `X-ray underluminous'. We chose to compare the X-ray luminosity of each group to its total $r$-band optical luminosity. The total group optical luminosity has been corrected to account for missing flux below the GAMA survey sensitivity limit and has been shown to correlate well with the group halo mass, as derived, for example, by weak lensing \citep[][]{Han:2015aa}. Using the total optical luminosity also avoids uncertainties introduced when using the velocity dispersion to calculate the group dynamical mass. We therefore use the L$^{XXL}_{300kpc}$ - L$_{opt}$ diagram to provide a measure of how over- or underluminous the X-ray emission from a group is in relation to its mass. However, we find that alternative methods, such as using group dynamical mass, velocity dispersion, or group multiplicity, yield similar results. 

\begin{figure}
\begin{center}
\includegraphics[width=\columnwidth]{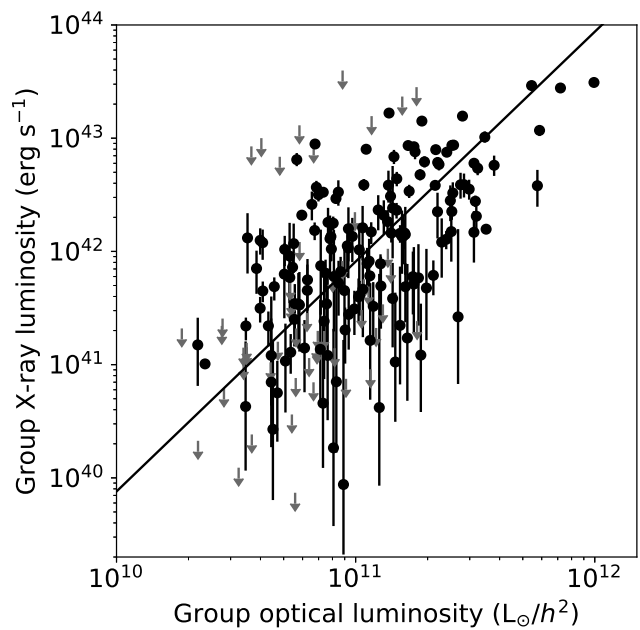}
\caption{L$^{XXL}_{300kpc}$ versus L$_{opt}$ for all GAMA groups, with $1\sigma$ uncertainties shown. The $90\%$ confidence upper limits on the X-ray luminosity are denoted with grey arrows. Five groups in which the GAMA centre is substantially offset from the X-ray position have been readjusted to reflect the aperture luminosity centred on the X-ray source position. The best fit line, using a bisection of linear fits, is shown as a black line.}
\label{fig:LxLopt_all}
\end{center}
\end{figure}

Figure \ref{fig:LxLopt_all} shows the L$^{XXL}_{300kpc}$ versus L$_{opt}$ diagram for all 201 groups in our sample. This includes the corrected luminosities for groups that have had their X-ray value changed because of a misalignment between the optical and X-ray centre (discussed in Sect. \ref{section:centre}).

We fit a straight line in order to separate the groups into X-ray overluminous and X-ray underluminous classes. We calculate a bisector of the linear fits $X|Y$ and $Y|X$, so as not to assume an independent variable. For this fit we only consider the detected sources, and not the upper limits. This best fit line is shown as a solid line in Fig. \ref{fig:LxLopt_all}. 

The gradient of our fit is $\frac{d(\log(L^{XXL}_{300kpc}))}{d(\log(L_{opt}))}=2.03$. This is somewhat lower than the value reported in \citet[][]{Wang:2011aa}, but agrees well with the slope of 2.05 derived by \citet{Osmond:2004aa} for a sample of galaxy groups with ROSAT X-ray observations, and agrees with theoretical expectations \citep[e.g.][]{Donahue:2001aa}. We note that as we are interested in the origin of the scatter in L$^{XXL}_{300kpc}$ versus L$_{opt}$ in this study, we use this simple linear fit to separate galaxies with different X-ray luminosities (see below). Our fitted gradient is used as the basis for our selection throughout the study. 

The metric we adopt as a measure of X-ray overluminosity is $L^{XXL}_{300kpc}/L_{opt}^{2.03}$, which we denote R$_{XO}$. The value of this quantity defines the position of a group relative to the fit line in Fig. \ref{fig:LxLopt_all}. 

\subsection{X-ray overluminosity sub-samples}
\label{Section:subsamples}
We divide our group sample into three different sub-samples based on the value of R$_{XO}$. Groups with R$_{XO}$ at least double the value corresponding to the best fit relation in Fig. \ref{fig:LxLopt_all} are considered X-ray overluminous, while groups with R$_{XO}$ at least three times lower than the best fit value are denoted X-ray underluminous. Finally, all groups lying between these values are `X-ray normal'. These values were chosen to provide reasonable sub-sample sizes of X-ray over- and X-ray underluminous groups. We note that the median percentage errors on our X-ray luminosities are $\sim40\%$, so statistical scatter can move only a modest fraction of the points from one sub-sample to another, as discussed in Sect. \ref{Section:biases}, where we also consider the impact of changing the threshold lines that separate the sub-samples.

\begin{figure}
\begin{center}
\includegraphics[width=\columnwidth]{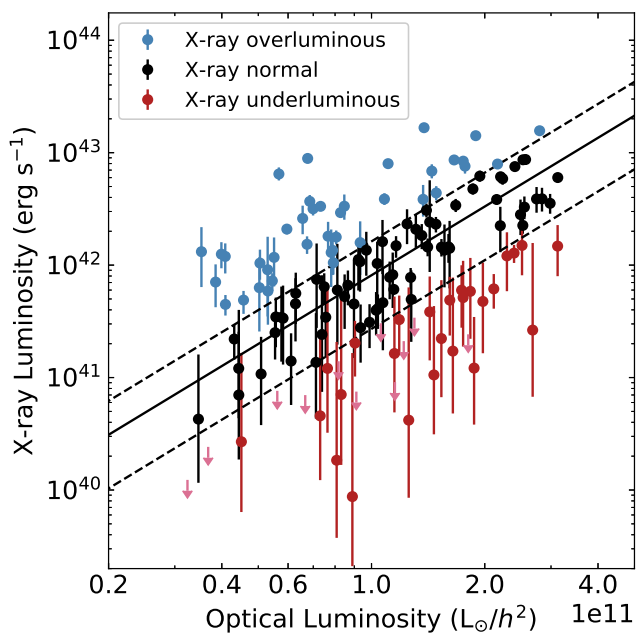}
\caption{L$^{XXL}_{300kpc}$ versus L$_{opt}$ for all groups in our final sample, with $1\sigma$ uncertainties shown. The trend line from Fig. \ref{fig:LxLopt_all} is shown as a solid black line. The dashed lines show the thresholds used in creating our X-ray overluminous and X-ray underluminous samples. These lines are a factor of 2 above and 3 below the best fit line, respectively. These values were chosen to achieve comparable numbers of groups in the bright and dim samples. More discussion on different choices for this selection can be found in Sect. \ref{Section:Discussion}.}
\label{fig:Xraycuts_final}
\end{center}
\end{figure}

While most of the groups with X-ray upper limits lie in the X-ray underluminous region, there are several sources with upper limits lying in the `normal' and `overluminous' regions. The upper limits contain important information about some of the most X-ray underluminous groups, but have to be used with care to avoid adding noise to our results. These sources pose a problem, as their unknown true value of $L_{X300kpc}$ could lie in any of the sub-samples below the recorded upper limit value. To avoid the danger of attributing them to the incorrect sub-sample and adding extra noise to our results, we remove all upper limit sources from both the normal and overluminous sub-samples for our analysis. However, groups with X-ray upper limits that fall in the underluminous region of Fig.\ \ref{fig:Xraycuts_final} are known to be underluminous, so these groups are retained in the X-ray underluminous sub-sample and are marked in Fig.\ \ref{fig:Xraycuts_final} with red down-arrows.

Ideally, our three group sub-samples would have similar distributions in redshift and in optical luminosity, in case either of these parameters has an impact on other group properties. To help achieve greater similarity between the distributions of these two properties, we truncate our sample by excluding groups outside the redshift range $0.05 < z < 0.35$ and the optical luminosity range $10.5 <$ log(L$_{opt}/$L$_{\odot}/h^{2}$) $< 11.5$. As an indication of the expected total mass of these groups, we can use the scaling relation between total Sloan Digital Sky Survey (SDSS) $r$-band luminosity and virial mass found by weak lensing mass from the work of \citet{Mulroy2018}. For our optical luminosity range, this scaling relation finds that the virial mass range is: $13.4 <$ log(M$_{group}/$M$_{\odot}$) $< 14.2$. 
The resultant sample of groups, split into respective X-ray overluminous, normal, and underluminous sub-samples is shown in Fig. \ref{fig:Xraycuts_final}.

After imposing group luminosity limits and redshift limits and removing upper limit points in the X-ray overluminous and X-ray normal samples, we are left with a final group sample of 142 groups, with 40, 65, and 37 X-ray overluminous, normal, and underluminous groups, respectively. These groups contain a total of 1163 galaxies, with 295, 538, and 330 galaxies in each of the respective sub-samples.

\begin{table}
\begin{center}
\begin{tabular}{ m{0.78\columnwidth} m{0.12\columnwidth} }
\hline
Sample & No.\ of GAMA Groups \\
\hline
Groups within XXL footprint & 238 \\  
High background XXL fields excluded & 232 \\
Groups with $>50\%$ aperture masking and poor X-ray photometry excluded & 213 \\
Groups near potential substructure excluded & 200 \\
Groups with $10.5 <$ log(L$_{opt}$) $< 11.5$ and $0.05 <$ z $ < 0.35$ & 175\\
Removal of upper limit groups from X-ray overluminous and normal subsamples (main sample used throughout) & 142 \\
Groups with  $0.05 <$ z $< 0.3$ (used in  Sect. \ref{Section:Galaxyproperties}) & 128 \\

\hline

\end{tabular}
\end{center}
\caption{Breakdown of the number of galaxy groups used in this study.}
\label{table:Xray_samples}
\end{table}

\subsection{Sub-sample comparisons}
\label{Section:comparisons}

Throughout this paper, we test for differences between sub-samples using the Anderson-Darling (A-D) test for k-samples as implemented in \citet{Scholz:1987aa}. In comparison to the commonly used Kolmogorov-Smirnov (K-S) test, the A-D test is more sensitive to a difference at the tail-end of the distributions, which is especially important in some of the distributions we examine. For consistency, we use the A-D test throughout the paper, but we note that qualitative similarities exist between the results from A-D and K-S tests. 

\begin{figure}
\begin{center}
\includegraphics[width=\columnwidth]{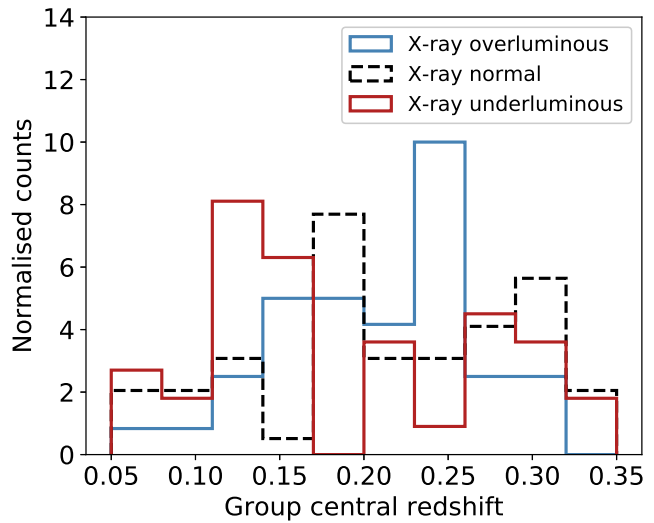}
\caption{Histogram of the group redshift for the X-ray overluminous (blue), X-ray normal (black dashed), and X-ray underluminous (red) sub-samples.
}
\label{fig:Zhist}
\end{center}
\end{figure}

Having applied the cuts to better account for differences between the overluminous and underluminous groups, we test the group sub-samples for differences in several group parameters. Figure \ref{fig:Zhist} shows the histogram of group central redshift for each sub-sample. We find the mean redshift of the three sub-samples are all similar (overluminous: $\bar{z}=0.21$, normal: $\bar{z}=0.23$ , underluminous: $\bar{z}=0.19$), so it is unlikely that redshift-dependent effects could introduce any significant biases into our results We note that using the A-D test to compare the redshift distributions between the X-ray overluminous and underluminous sub-samples and find a difference between these sub-samples has a probability of chance occurrence of $p\approx0.036$. This constitutes a $>2\sigma$ result, and as such may be a selection effect. However, it is unlikely for this difference, despite being statistically significant, to account for any differences in group properties in the following sections.

\begin{figure}
\begin{center}
\includegraphics[width=\columnwidth]{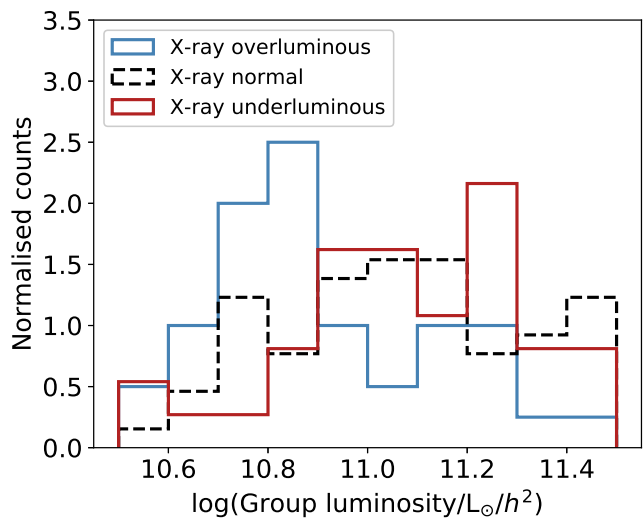}
\caption{Histogram showing the distribution of group log(L$_{opt}$) for X-ray overluminous (blue), X-ray normal (dashed black), and X-ray underluminous (red) sub-samples. We find the average group L$_{opt}$ increases from X-ray overluminous to X-ray underluminous samples.}
\label{fig:Lopthist}
\end{center}
\end{figure}

Additionally, despite the limits imposed on L$_{opt}$, the sub-samples still have significant differences in their L$_{opt}$ distributions. Figure \ref{fig:Lopthist} shows the distributions of log(L$_{opt}$) for the three sub-samples. An A-D test comparing the X-ray overluminous and underluminous sub-samples confirms a very significant ($p = 0.0013$) difference. We find that the X-ray underluminous sample has a mean optical luminosity of $12.1\times10^{10}$L$_{\odot}/h^{2}$, $\sim1.5$ times higher than the X-ray overluminous sample value (L$_{opt}=8.1\times10^{10}$L$_{\odot}/h^{2}$).

This difference in L$_{opt}$ between the group sub-samples means that 
differences in properties between the sub-samples could result from differences in L$_{opt}$, as well as differences in X-ray status.
For example, previous studies have found evidence that groups with higher optical luminosity tend to have a higher fraction of red, evolved galaxies, and a larger magnitude gap between first and second ranked galaxies \citep[e.g.][]{Weinmann:2006aa, Davies:2019aa}. In the case of group properties, we are able to scale several of our parameters by their L$_{opt}$ values, in order to remove any expected dependence, as discussed below. Moreover, as will be seen in Sect. \ref{Section:Results}, the trends we find in galaxy properties across our sample run in the {opposite} direction to what would be expected if they were driven by differences in L$_{opt}$.

\section{Results}
\label{Section:Results}
We now compare the properties of the groups and their member galaxies across the three group sub-samples.

\subsection{Group structural parameters}
\label{Section:Structure}
For each group, a range of group properties are available from the GAMA G$^{3}$Cv10 catalogue, described briefly in Sect. \ref{Section:GAMA}. Further details of these group parameters and how they have been calculated are presented in \citet{Robotham:2011aa}.

In this study, we aim to test differences between the group structural parameters in each of the 3 sub-samples. We consider the group $50\%$ radius ($R_{50}$), axis ratio (B/A), internal velocity dispersion ($\sigma$), velocity skew, the offset of the brightest group galaxy (BGG) from the optical luminosity weighted centre, and the $r$-band magnitude gap between brightest and second brightest member galaxies for all groups in each sub-sample. Additionally, we calculate scaled versions of group radius, dynamical mass (calculated using Eq. 18 in \citealt{Robotham:2011aa}), and velocity dispersion, to remove the expected dependence on L$_{opt}$ for self-similar systems with uniform mass-to-light ratios. As noted in Sect. \ref{Section:comparisons}, the distribution of L$_{opt}$ varies between our three group subsets. The scaled group $50\%$ radius, dynamical mass, and velocity dispersion are $R_{50}/L_{opt}^{1/3}$,
 $M/L_{opt}$, and $\sigma/L_{opt}^{1/3}$, respectively. We also calculated a group crossing time based on the study of \citet{Ai:2018aa}, defined as
\begin{equation}
  t_{cross} = \frac{1.51^{1/2}R_{50}}{3^{1/2}\sigma} .
\end{equation}
For each parameter we compare the mean values and distributions of each sub-sample, to determine if there are any significant structural differences between the group subsets. The two-sample A-D test is used to compare the distributions of each parameter between the X-ray overluminous and X-ray underluminous sub-samples.

\begin{figure}
\begin{center}
\includegraphics[width=\columnwidth]{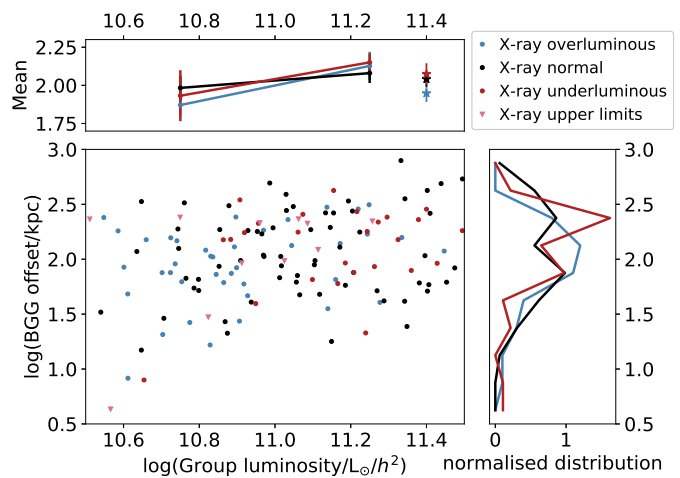}
\caption{Values of the offset of the central galaxy from the optical luminosity weighted centre for each member of the X-ray overluminous (blue), X-ray normal (black), and X-ray underluminous (red) sub-samples. Underluminous groups with X-ray upper limits are denoted with a light red triangle. The main panel shows the values for each group, while the right panel shows the distribution of  central galaxy offset for each of the sub-samples. The top panel shows the sub-sample means, both across the full range in L$_{opt}$ (stars) and split into two bins in L$_{opt}$ (circles). X-ray underluminous systems have, on average, larger offsets than the X-ray overluminous groups. 
}
\label{fig:BCGoffset_3panel}
\end{center}
\end{figure}

Looking first at the BGGs, we find that groups in the overluminous sub-sample have a smaller offset between the BGG and the optical luminosity weighted group centre, compared with the X-ray underluminous sub-sample. Figure \ref{fig:BCGoffset_3panel} shows the log(BGG offset) for all groups classed as X-ray overluminous (blue), X-ray underluminous (red), and X-ray normal (black). We see in the upper panel the means for each category, showing that the X-ray underluminous groups have a higher BGG offset in each of two luminosity bins (points) and when averaged across the whole sample (stars). The median offset of the BGG in the overluminous sample is 98 kpc, $\sim1.8$ times smaller than in the X-ray underluminous sample (175 kpc). The right hand panel of Fig. \ref{fig:BCGoffset_3panel} compares the distribution of each sample, which highlights the excess of high offset systems in the X-ray underluminous sample, compared with the overluminous sample. An A-D test confirms that these two distributions differ with a chance probability $p<0.04$.

\begin{figure}
\begin{center}
\includegraphics[width=\columnwidth]{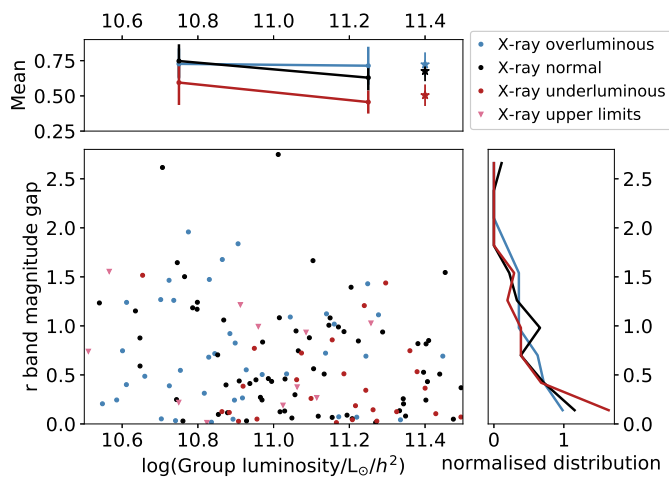}
\caption{Similar to Fig. \ref{fig:BCGoffset_3panel}, showing the difference in magnitude between the two brightest member galaxies of each group. We see a trend in magnitude gap from the X-ray underluminous groups to the overluminous groups.}
\label{fig:maggap_3panel}
\end{center}
\end{figure}

Secondly, we see that the magnitude gap -- the difference in magnitude between the brightest and second brightest group member galaxies -- shows a difference across the group subsets. Figure \ref{fig:maggap_3panel} displays the magnitude gap for all groups, in a similar format to Fig. \ref{fig:BCGoffset_3panel}. The magnitude gap in X-ray overluminous groups is on average 0.22 mag higher than for the X-ray underluminous sub-sample. It can be seen from the rightmost panel in Fig. \ref{fig:maggap_3panel} that the difference between the distributions is concentrated at the ends of the distribution -- in particular, many of the X-ray underluminous groups have very small magnitude gaps. Despite this, an A-D test does not show a very significant difference between the distribution of magnitude gap in the sub-samples, with a $p$ value of $\approx 0.1$. We discuss this further in Sect. \ref{Section:high_nfof}.

We also used the A-D test to search for significant differences between the X-ray overluminous and underluminous sub-samples in other group structural properties: axis ratio, velocity skew, scaled velocity dispersion, scaled group mass, scaled group radius, and crossing time. None of these tests showed discrepancies at the $2\sigma$ level. 

\begin{figure}
\begin{center}
\includegraphics[width=\columnwidth]{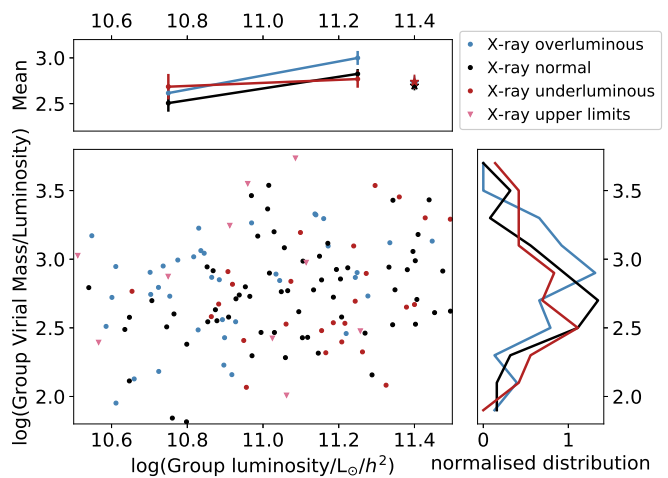}
\includegraphics[width=\columnwidth]{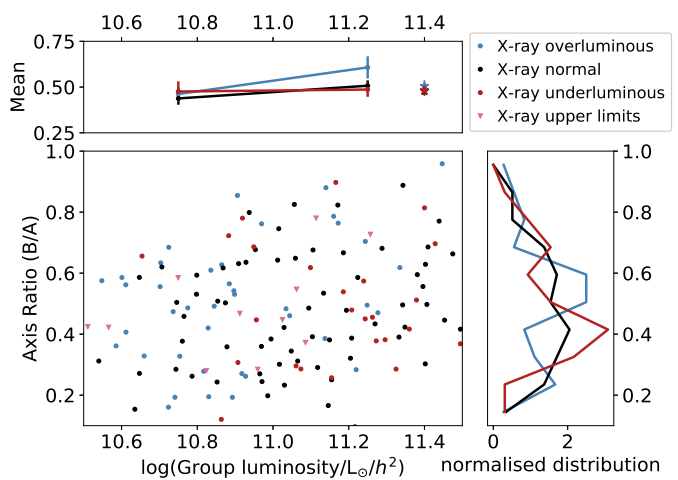}
\caption{Similar to Fig. \ref{fig:BCGoffset_3panel}, showing the difference in the group scaled mass ratio (Group Mass/L$_{opt}$, top) and group axial ratio (B/A, bottom). In both plots there appears to be a difference between the over- and underluminous groups in high L$_{opt}$ systems.  This is explored further in Sect. \ref{Section:high_nfof}. However, a comparison of the full distribution for either parameter does not show significant differences between the overluminous and underluminous sample.}
\label{fig:massrat_3panel}
\end{center}
\end{figure}

However, examining the distribution plots analogous to Fig. \ref{fig:maggap_3panel}, we see evidence of systematic differences in a number of the structural properties between the three X-ray luminosity classes in the higher of the two L$_{opt}$ bins. For example, Fig. \ref{fig:massrat_3panel} shows these plots for the scaled group mass (effectively the virial mass-to-light ratio) and the axis ratio. While the full distributions of these parameters are not statistically distinguishable, we do see differences in the mean values for sources with log(L$_{opt}) > 11$ (the higher of the two L$_{opt}$ bins in the top panel).

In both cases, the mean values of the structure parameters are ordered such that overluminous $>$ normal $>$ underluminous, with deviations that are larger than the uncertainties. We compare the mean values of the overluminous and underluminous groups with log(L$_{opt}) > 11$. A two-sided t-test of means gives a significance of $p=0.067$ and $p=0.11$ when comparing the difference in means in the scaled mass and axial ratio between the underluminous and overluminous sources, respectively. While a comparison of the magnitude gap and the mass ratio for the full underluminous and overluminous sub-samples did not show any statistically significant difference, the differences in mean values seen in high L$_{opt}$ groups may suggest that larger groups do have some structural differences.

Assuming, for the moment, that these are real effects, why might they be visible only in the higher of the two optical luminosity bins? It is important to bear in mind that, despite the power of the GAMA survey, the majority of our groups contain only 5-7 catalogued galaxies. With such small numbers, the stochastic noise in calculating structural parameters is large. For example, the velocity dispersion will be subject to a sampling error of over $\sim35\%$ if only five group members are available from which to calculate it \citep[][]{Ruel:2014aa}.

This sampling noise will dilute the significance of any real trends in the data.
Since the groups with high L$_{opt}$ also tend to have higher multiplicity (the mean multiplicity of groups with L$_{opt}$>$10^{11}L_{\odot}$/$h^{2}$ is 9.8, whilst for those with L$_{opt}\le10^{11}L_{\odot}/h^{2}$ it is 6.3) they will be subject to less sampling noise, making any trends in the structural parameters more apparent. We explore this idea further in Sect. \ref{Section:high_nfof}.

\subsection{Galaxy properties}
\label{Section:Galaxyproperties}
We now investigate differences in the $(g - i)$ colour and the H$\alpha$-based star formation rate of the member galaxies within our three group subsets. The colour and star formation properties of galaxies are known to depend on their absolute magnitude or stellar mass, so we must restrict the examination to a stellar mass or magnitude limit that can be consistently probed for all groups regardless of their redshift.

GAMA is an $r$-band selected survey, so the clearest selection is an $r$ absolute magnitude limit. For the remainder of the paper, we only consider group members brighter than an $r$-band absolute magnitude of M$_{r} = -21.1$ mag, within a redshift range of 0.05 $<$ z $<$ 0.3. This leaves a sample of 128 groups, which we investigated (see Table \ref{table:Xray_samples} for details). We could have chosen a fainter (brighter) magnitude limit, which would have led to a smaller (larger) maximum redshift, but we obtain qualitatively similar results for magnitude limits varied between magnitudes M$_{r} = -20$ and $-21.1$ mag.

We calculated absolute magnitudes and colours for all galaxies with GAMA spectral quality nQ $>2$, using photometric values derived from the SDSS data release 8 photometry \citep{Aihara:2011aa}. Absolute magnitudes and optical colours are calculated, correcting for galactic dust using maps from \citet{Schlegel:1998aa}, and K-corrected to redshift $z=0$ using the analytical approximations of \citet{Chilingarian:2010aa}. 
We estimated the stellar masses of galaxies using the relation of \citet{Taylor:2011aa}, which uses the absolute i-band magnitude and the (g - i) colour:
\begin{equation}
  log(M_{stellar}) = 1.15 + 0.70 \times (g - i) - 0.4 \times M_{I}  
.\end{equation}
We calculated this for each of the $\sim7700$ galaxies in the magnitude limited sample. For group members in our sample brighter than an $r$-band absolute magnitude of M$_{r} = -21.1$ mag, we find the mean galaxy stellar mass of the ten faintest galaxies to be log(M$_{stellar}/$M$_{\odot}) \approx 10.10$.

\subsubsection{Colour}
\label{Section:colour}
Colours of galaxies are well known to be bimodal, with the two populations commonly denoted as `red' or `blue' \citep[e.g.][]{Strateva:2001aa, Baldry:2004aa, Balogh:2004aa}. We compare the ratio of the blue population to the total population in each of the three sub-samples. We split the sample by fitting the $(g - i)$ colour distributions of galaxies in separate mass bins with two separate Gaussian curves (similar to that used in \citealt{Crossett:2017aa}, but see also \citealt{Taylor:2011aa} and references therein). The intersection of the two Gaussian curves was then taken in each mass bin, and the resultant values were used to construct a line to separate red from blue galaxies. The equation of the derived line is
\begin{equation}
  (g - i) = 0.064\times \textrm{log}(M_{stellar}) + 0.329
.\end{equation}

\begin{figure}
\begin{center}
\includegraphics[width=\columnwidth]{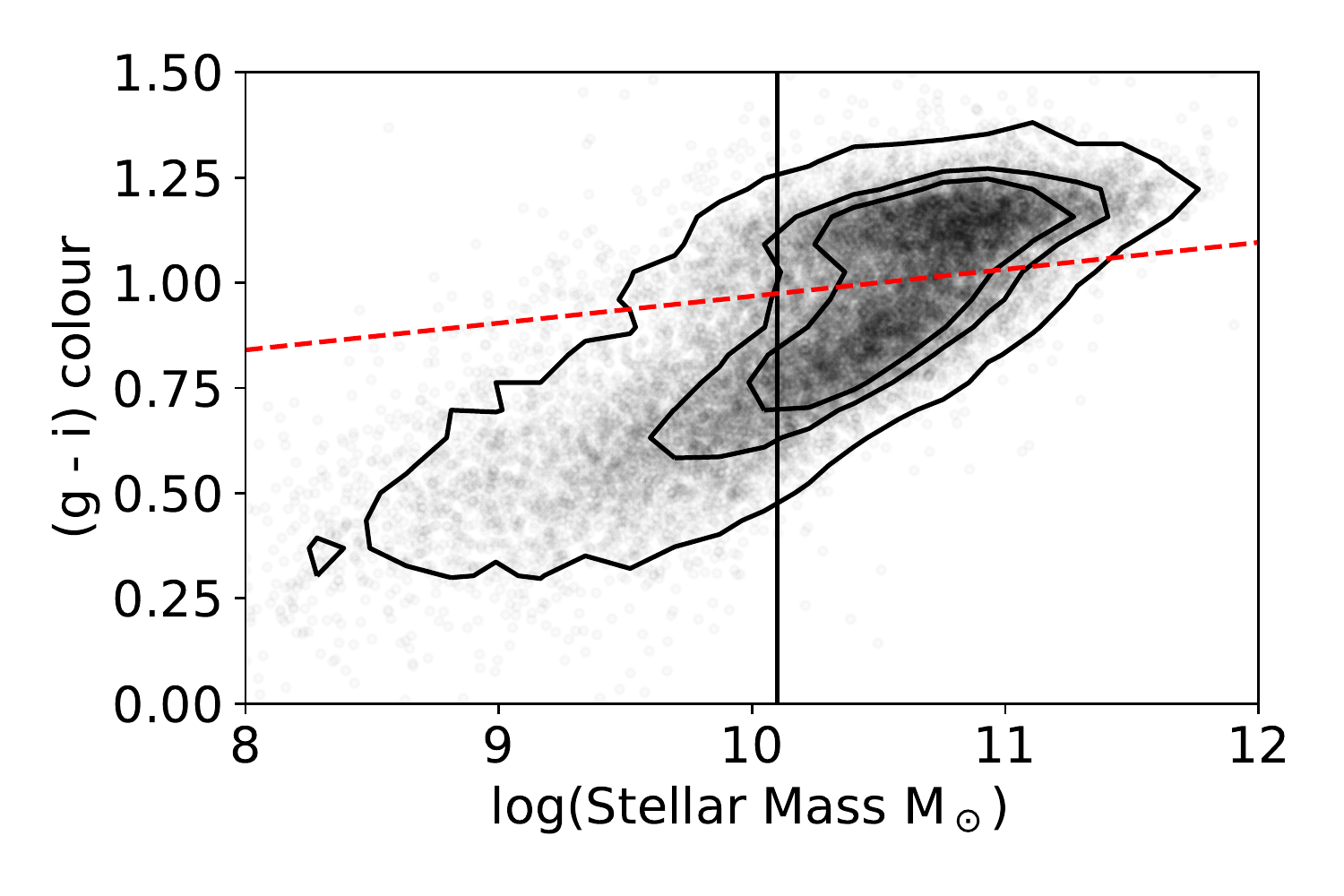}
\caption{(g - i) colour versus stellar mass diagram for all galaxies in the GAMA G02-XXL overlap region. The red line denotes a linear fit calculated by fitting Gaussian curves to the red and blue populations. We consider galaxies as blue if they lie below the linear fit, and red if they lie above it. The solid black line denotes the mean mass of the ten faintest galaxies to our M$_{r} = -21.1$ magnitude limit, corresponding to log(M$_{stellar}/$M$_{\odot}) \approx 10.10$.}
\label{fig:colour_mass_diagram}
\end{center}
\end{figure}

Figure \ref{fig:colour_mass_diagram} shows all galaxies within the GAMA G02 - XXL overlap region, including galaxies not affiliated with any group. The line shown divides the sample into the blue and red populations. 

\begin{figure}
\begin{center}
\includegraphics[width=\columnwidth]{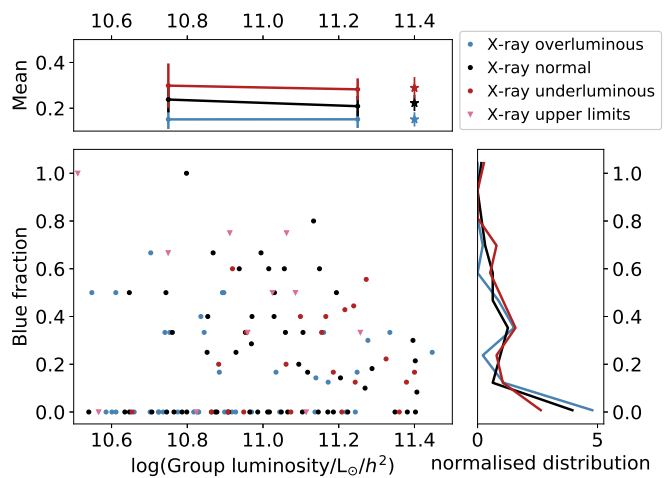}
\caption{Similar to Fig. \ref{fig:BCGoffset_3panel}, showing the fraction of blue galaxies in each group. The blue fraction is calculated for galaxies brighter than our absolute magnitude limit of M$_{r} = -21.1$ mag. We find that the X-ray overluminous groups have a lower blue fraction across the entire L$_{opt}$ range compared with the X-ray underluminous groups. Galaxies in X-ray `normal' groups are intermediate.}
\label{fig:redfrac_3panel}
\end{center}
\end{figure}

We then use this split in colour to calculate the fraction of blue galaxies within each group brighter than our M$_{r} = -21.1$ mag limit. Figure \ref{fig:redfrac_3panel} shows the blue fraction for each group in the sample, with colours representing the X-ray overluminous, underluminous, and normal sub-samples. As in Fig. \ref{fig:maggap_3panel}, the mean values are shown in the upper panel, and the distribution is displayed in the right panel. Many of the groups here have blue fractions equal to zero, indicating an absence of blue galaxies in the group above the limiting magnitude. We see that the overluminous groups have a lower blue fraction on average when compared with the X-ray underluminous and normal groups.  The rightmost panel in the figure shows that the main deviation in the distribution lies at the bottom end, where a large fraction of X-ray luminous groups have zero blue fraction. The A-D test shows a clear difference between the under- and overluminous samples ($p$ = 0.0074). The lower blue fraction in X-ray overluminous groups is similar to that seen in large clusters in \citet{Wang:2014aa}. However, our results are seen in group-sized haloes, highlighting that this result may hold over a large range of halo mass.

\subsubsection{Star formation rate}
In addition to the colour of galaxies in each of the group sub-samples, we also compared the star formation rate of each sub-sample using H$\alpha$ measurements from \citet{Gordon:2017aa}. We calculated the H$\alpha$ luminosity, L$_{H\alpha}$ using

\begin{multline}
L_{H\alpha} = (EW_{H\alpha} + EW_{c}) \times 10^{-0.4 \times (M_{r}-34.1)} \\ \times \frac{3 \times 10^{18}}{(6564.61 \times (1+z))^{2}} \times \left(\frac{F_{H\alpha}/F_{H\beta}}{2.86}\right)^{2.36} ,
\end{multline}

\noindent where EW$_{H\alpha}$ is the equivalent width of H$\alpha$, EW$_{c}$ is the correction to H$\alpha$ to account for emission filling (taken as 2.5 from \citealt{Hopkins:2013aa}), and the $\frac{F_{H\alpha}}{F_{H\beta}}$ is the flux ratio of H$\alpha$ to H$\beta$ emission, used to correct for dust obscuration in each galaxy. This equation has been used numerous times to calculate H$\alpha$ luminosities \citep[e.g.][]{Gunawardhana:2011aa, Hopkins:2013aa, Davies:2016aa}.

\begin{figure}
\begin{center}
\includegraphics[width=\columnwidth]{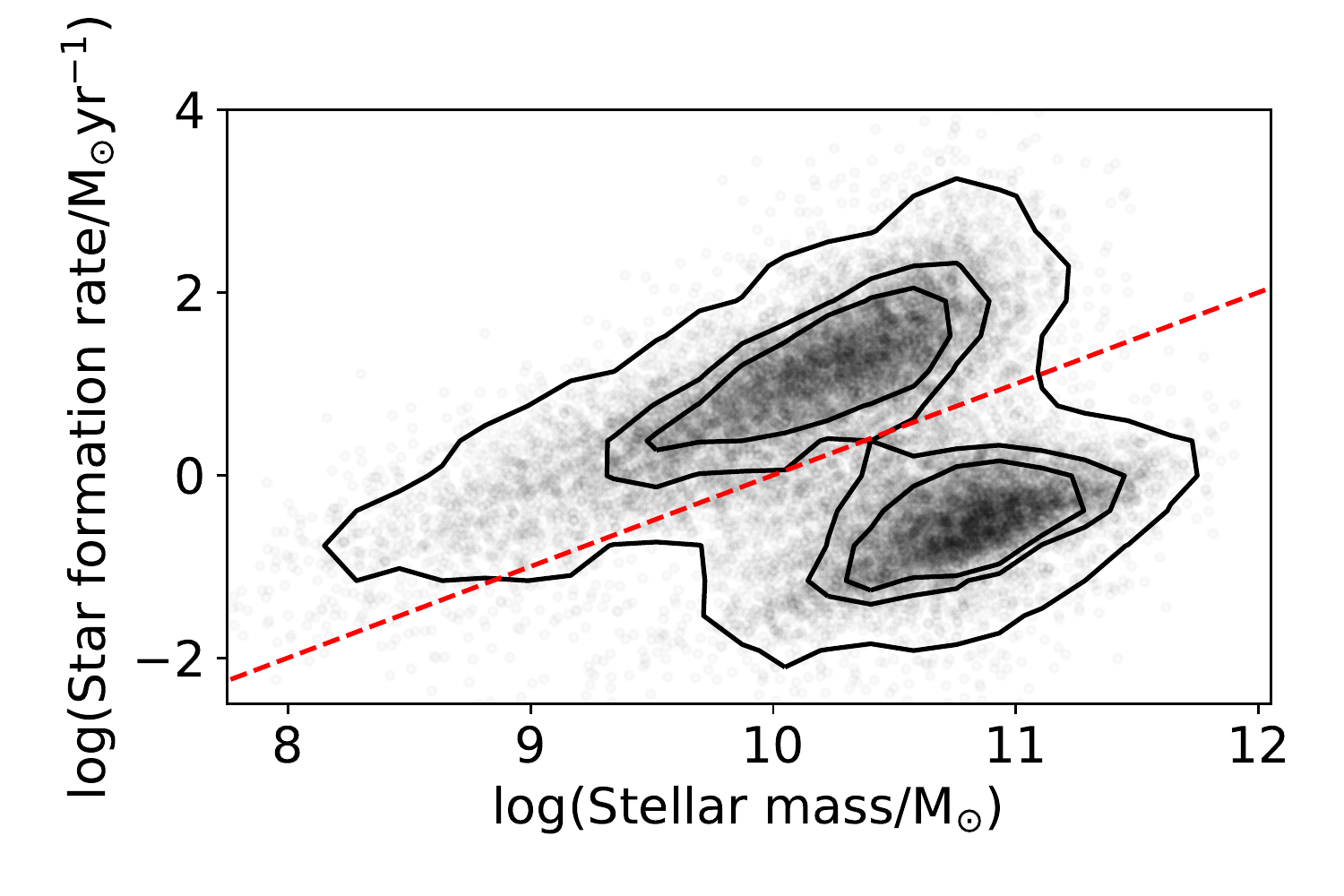}
\caption{Star formation rate versus stellar mass for all galaxies in the GAMA G02-XXL overlap region. A main star forming sequence and a passive cloud are clearly distinguished. We separate the star forming galaxies and passive galaxies using a constant specific star formation rate of log(sSFR) = -10 yr$^{-1}$ (dashed red line).}
\label{fig:SF_mass_diagram}
\end{center}
\end{figure}

We then converted this to a star formation rate using the relation given in \citet{Davies:2016aa},\begin{equation}
SFR_{H\alpha} = \frac{L_{H\alpha}}{1.27\times 10^{34}}\times1.53.
\end{equation}
The star formation rates of all galaxies in the G02-XXL overlap region are shown in Fig. \ref{fig:SF_mass_diagram}, plotted against stellar mass. 
We separate the sample into star forming and passive galaxies using the specific star formation rate (sSFR; the star formation rate divided by stellar mass), with a threshold value set as log(sSFR) = -10 yr$^{-1}$. This relation is shown in Fig. \ref{fig:SF_mass_diagram}, and clearly separates the main populations. We use this separation below, but find similar results using a relation based on that of \citet{Davies:2016aa}.

\begin{figure}
\begin{center}
\includegraphics[width=\columnwidth]{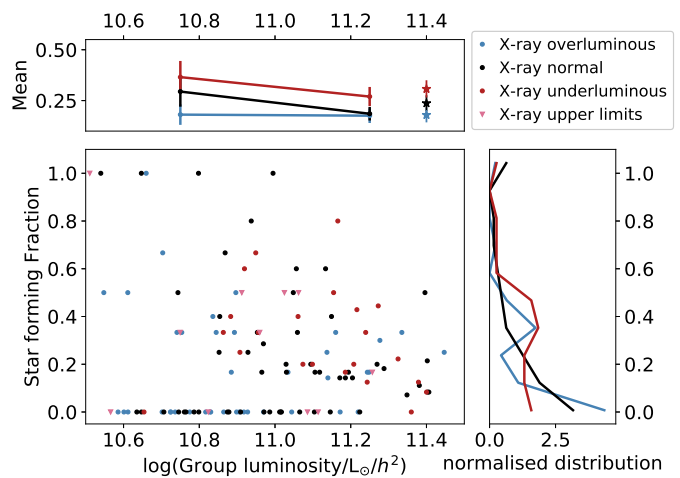}
\caption{Same as Fig. \ref{fig:maggap_3panel}, showing the fraction of star forming galaxies brighter than M$_{r} = -21.1$ mag in each group. The X-ray overluminous groups have a low star forming fraction across the entire L$_{opt}$ range, while star formation in the underluminous groups depends more on the L$_{opt}$ value.}
\label{fig:SFfrac_3panel}
\end{center}
\end{figure}

Figure \ref{fig:SFfrac_3panel} shows the fraction of star forming galaxies for all groups in our sample, in the same format as Fig. \ref{fig:redfrac_3panel}. This shows that the star forming fraction in X-ray overluminous groups is low across the entire L$_{opt}$ range, whereas the X-ray underluminous groups have a star forming fraction that more strongly depends on L$_{opt}$. Over the whole L$_{opt}$ range, the fraction of star forming galaxies in the X-ray underluminous groups is on average 1.7 times higher than that of the overluminous groups, and an A-D test finds that the distributions of the over- and underluminous sub-samples differ with $p = 0.0023$. The relationship we find between X-ray overluminosity and star formation fraction matches results from \citet{Roberts:2016aa}, who found a lower star forming fraction in X-ray bright SDSS groups and clusters. Our results show that this link between X-ray overluminosity and galaxy star formation extends to some of the most X-ray underluminous and low mass systems.

Finally, we check to see if the three sub-samples have a difference in the fraction of active galactic nuclei (AGN), which might impact on our star formation results, since AGN can also generate high H$\alpha$ luminosity. We compare the fraction of galaxies that host an AGN in the three sub-samples, using the emission line diagnostic of \citet{Baldwin:1981aa}. We find that the AGN fraction, using either the emission line definitions of \citet{Kewley:2001aa}  or \citet{Kauffmann:2003aa}, does not differ significantly between the three X-ray sub-samples and is in any case modest; $10\%$ of our galaxies are classified as AGN in the definition of \citet{Kewley:2001aa}. We therefore conclude that the differences in the star forming fraction that we identify cannot be driven by AGN emission. A full analysis of the incidence of AGN in these three sub-samples, including the analysis of both optical and X-ray AGN is beyond the scope of the present work, but may be investigated in more detail in future studies of these groups.

\section{Discussion}
\label{Section:Discussion}

\subsection{X-ray luminosity as a measure of group age}

The results discussed above demonstrate that a number of group properties show systematic trends across the sequence: X-ray underluminous, through X-ray normal, to X-ray overluminous systems. Increasing X-ray overluminosity is associated with a decrease in blue galaxies, a reduction in star formation and increasing dominance of the brightest member galaxy (found by both the brightest galaxy offset from the optical centre and by the gap in $r$-band magnitude), as well as an increase (only detected at a significant level in the high multiplicity systems, as reported in Sect. \ref{Section:high_nfof}) in the virial mass-to-light ratio. These trends cannot be driven by the differences in mean optical luminosity between our three group subsets discussed in Sect. \ref{Section:comparisons}, since most go against generally observed trends with L$_{opt}$. We recall that our X-ray underluminous groups have {higher} average group L$_{opt}$. In general, cluster richness is found to correlate positively with properties such as the passive fraction, and magnitude gap \citep[e.g.][]{Davies:2019aa}, whilst we find that our X-ray underluminous sub-sample shows {low} passive fraction and magnitude gap, despite having, on average, higher L$_{opt}$ than the other two sub-samples.

These trends can be understood on the basis of group evolution. As groups evolve, they are expected to deepen their gravitational potential well, compress intergalactic gas, and develop a dense hot IGM, which emits X-rays. The X-ray emissivity is proportional to the square of the gas density, which will therefore increase during this process.

Similarly, as the group evolves, the galaxy populations will quench their star formation and evolve to become passive and redder. At the same time, galaxy merging within the low velocity dispersion group environment can build a dominant central galaxy residing at the bottom of the gravitational potential. Since the more massive galaxies lose orbital energy and merge preferentially, this process also increases the magnitude difference between the brightest and second brightest group members \citep[e.g.][]{Smith:2010aa, Oliva-Altamirano:2014aa, Raouf:2019aa}.

We therefore suggest that the position of a group in the L$^{XXL}_{300kpc}$ versus L$_{opt}$ diagram is a useful measure of group evolutionary state. Dynamical evolution of a group will cause it to move vertically upwards in the diagram, as its IGM is compressed, increasing its temperature and density, and hence its X-ray luminosity. Growth of a virialised system through accretion and merging will move systems diagonally upwards, from bottom left to top right, as both optical and X-ray luminosity rise.

Of course, even if dynamical evolution plays a dominant role in determining X-ray properties, it is still possible that other factors are also at play. We explore this in more detail in Sect. \ref{Section:EXdim}.

\subsection{Group structural parameters in larger systems}
\label{Section:high_nfof}
Given the interpretation above, it seems surprising that we do not see more significant differences in group structural properties. For example, a group in the early stages of formation might be expected to have a larger radius, lower velocity dispersion, and higher axis ratio and velocity skew. Three factors might make it difficult to detect such effects. Firstly, the evolution of structural properties as a group evolves is not necessarily simple and monotonic. For example, even for a spherically symmetric collapse the velocity dispersion will increase as a group starts to form, reach a maximum as the system passes through maximum collapse and then settle down at an intermediate value as it virialises. Hierarchical merging and departures from spherical symmetry add further complexity. 

Secondly, the visibility of some structural changes may depend upon viewing direction (for example, for axis ratio and velocity skew). Thirdly,  structural properties will be subject to large stochastic fluctuations for systems with only a few member galaxies. The trends in some structural parameters discussed in Sect. \ref{Section:Structure}, which are seen only in the high L$_{opt}$ groups, suggest that this last factor may be at work.

Our large sample of groups may help offset any effects of the first two factors, although we cannot completely rule these out. However, it is possible to mitigate the effects of the third factor. To explore the possibility that trends in structural properties are being lost in the statistical noise resulting from groups with low numbers of catalogued galaxies, we repeated our analysis including only groups with at least eight galaxy members. This further reduces the sub-samples to 11, 36, and 16 groups in the X-ray overluminous, normal, and underluminous sub-samples, respectively. Of course, the reduction in the number of groups available will tend to weaken the sensitivity of our tests, but this may be offset by the increased reliability in the value of the structure parameters for each group included.

\begin{figure}
\begin{center}
\includegraphics[width=\columnwidth]{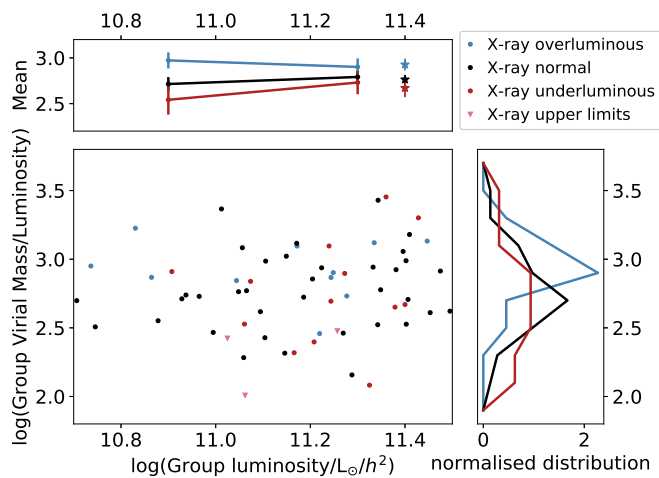}
\caption{Distribution in the scaled group virial mass for the three sub-samples when only high multiplicity groups ($N_{\rm gal}$>7) are included. X-ray overluminous groups show a higher scaled mass compared to X-ray underluminous systems.}
\label{fig:mass_cut_3panel}
\end{center}
\end{figure}

The most significant new result from this analysis of the high multiplicity groups is for the scaled group mass (defined as $M/L_{opt}$). As shown in Fig. \ref{fig:mass_cut_3panel}, when we only consider groups with at least eight members the group virial mass-to-light ratio is systematically larger in the overluminous groups compared with the underluminous sample, with the X-ray normal groups being intermediate. An A-D test shows the under- and overluminous samples to differ in their distribution with a significance of $p=0.038$; X-ray overluminous groups have a larger dynamical mass for their optical luminosity compared with the X-ray underluminous groups. 

As we discussed in Sect. \ref{Section:Structure}, there appeared to be a difference in the mean magnitude gap between X-ray overluminous and underluminous systems. Despite the visual difference, this was found not to be a statistically significant result for the full group sample, even at the $2\sigma$ level.
However, when we consider only the high multiplicity groups (with n$\geq8$), the A-D test does show a significant difference in the r-band magnitude gap between the overluminous and underluminous sub-samples ($p = 0.03$). This increased significance is consistent with Fig. \ref{fig:maggap_3panel}, which showed that the difference in the sub-sample magnitude gaps is larger at high group luminosity (top panel) and thus group multiplicity. It is plausible that the luminosity-dependent difference is a result of more massive groups having been virialised for a greater amount of time, and a resulting longer time to build up a significant BGG. However, none of the other structural parameters showed strong evidence of such a trend in the high multiplicity systems.

\subsection{Evolved X-ray underluminous groups}
\label{Section:EXdim}
While our principal result is that galaxy groups with different L$^{XXL}_{300kpc}$ versus L$_{opt}$ values appear to be separated by group evolutionary state, this does not explain all groups in our sample. There are several groups with predominantly old stellar populations that do not have significant X-ray emission. 

To explore the nature of this subset of groups, we identified groups from our X-ray underluminous sub-sample that have properties suggesting that they are old, evolved systems. We do this by requiring them to have a low star forming fraction and a short group crossing time. Additionally, we require that the $90\%$ upper limit to their L$^{XXL}_{300kpc}$ probability distribution (denoted L$_{UX300kpc}$) lie below the X-ray overluminosity threshold, L$_{UX300kpc}$/L$_{opt}^{2.03} = 2 \times 10^{19}$, separating the underluminous and normal group subsets. This helps ensure that these are genuinely X-ray underluminous systems rather than normal systems that have statistically scattered towards low L$^{XXL}_{300kpc}$.

\begin{figure}
\begin{center}
\includegraphics[width=\columnwidth]{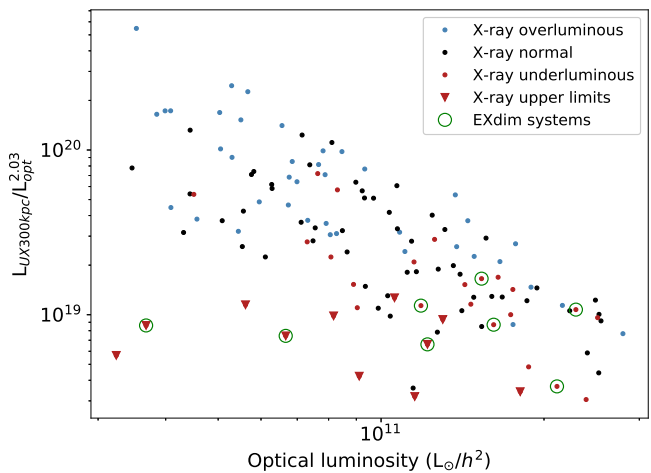}
\caption{Ratio of L$_{UX300kpc}$/L$_{opt}^{2.03}$ (U$_{XO}$) plotted against L$_{opt}$ for all groups. Blue, black, and red colours denote groups in the X-ray overluminous, normal and underluminous subsamples. Inverted triangles denote X-ray upper limits, where no significant X-ray flux value was detected. We highlight X-ray underluminous systems with properties suggesting an evolved group (called here EXdim groups) with green circles. If these groups are dynamically evolved their low X-ray emission must have some other cause.}
\label{fig:UX_EXdim}
\end{center}
\end{figure}

In Fig. \ref{fig:UX_EXdim} we plot the group L$_{UX300kpc}$/L$_{opt}^{2.03}$ ratio (hereafter U$_{XO}$) against L$_{opt}$ for all groups in our sample. We selected evolved, but X-ray underluminous (hereafter EXdim), groups to have a star forming fraction $< 0.3$, log(t$_{cross}) < -3.5,$ and L$_{UX300kpc}$/L$_{opt}^{2.03} < 2 \times 10^{19}$. There are eight such groups, constituting about 20\% of our sub-sample of X-ray underluminous groups.

If these EXdim systems are indeed evolved systems then they should contain hot gas. Their low X-ray luminosity then implies that this hot IGM must have low density, and hence low X-ray emissivity.
Two different explanations for such a low IGM density present themselves. Firstly, there could be a low gas fraction in these groups due to a high star formation efficiency \citep[e.g.][]{Hicks:2008aa} having converted most of the baryonic content of the group into stars. Alternatively, the entropy of the IGM may have been raised by feedback processes \citep[e.g.][]{Bower:1997ab, Hicks:2008aa, Pearson:2017aa}, resulting in a very extended hot gas distribution, with low gas density in the inner regions where most of the X-ray emission originates.

\begin{figure}
\begin{center}
\includegraphics[width=\columnwidth]{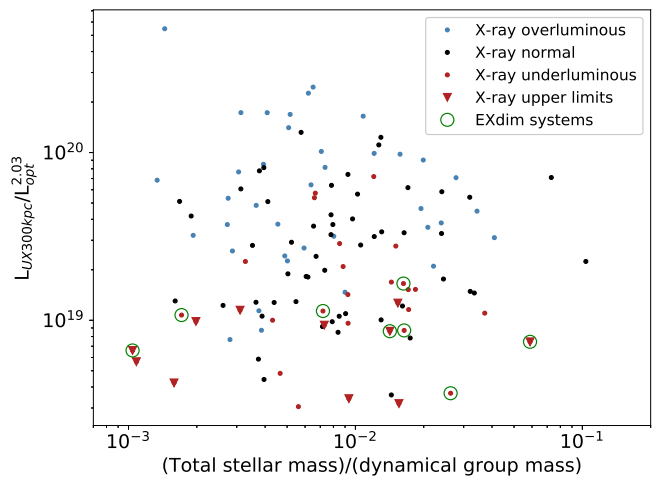}
\caption{U$_{XO}$ versus the group total group stellar mass/group 
dynamical mass. Symbols are as in Fig. \ref{fig:UX_EXdim}. The EXdim systems are shown with green circles. These evolved X-ray dim groups clearly do not preferentially fall towards the right of the diagram and hence do not have an unusually large fraction of their baryons locked into stars.}
\label{fig:UxMassrat}
\end{center}
\end{figure}

In order to test whether these EXdim groups are unusually efficient in forming stars compared to other groups, we compare the total mass of group member galaxies to the dynamical mass for each group. Figure \ref{fig:UxMassrat} plots the L$_{UX300kpc}$/L$_{opt}^{2.03}$ value against the total galaxy stellar mass/dynamical group mass for all groups in our sample. The groups are split into our three X-ray sub-samples, and the EXdim systems are highlighted.

We see that the EXdim systems occupy the full range of stellar fraction values, with no preference for high stellar mass/group dynamical mass. 
This indicates that efficient star formation is not the cause of the low X-ray emission in these evolved groups. We conclude that a high entropy IGM is a more likely candidate for the EXdim systems than a high star formation efficiency. However, deeper X-ray imaging would be required to detect any low entropy gas in order to further test this theory.

\subsection{X-ray versus optical selection}
\label{Section:Xrayoptical}
Our results demonstrate that the X-ray luminosity of galaxy groups of a 
given mass is primarily linked to their evolutionary state. Systems with low X-ray luminosity are more likely to be dynamically young, and hence incompletely virialised. However, approximately 20\% of our X-ray underluminous sub-sample appear to be mature systems, in which the low
X-ray luminosity is most likely the result of strong feedback having
raised the entropy of the IGM. This sub-sample of our groups, which we denote
our EXdim sample, amounts to $6\%$ of the total sample of 142 groups we have 
analysed in detail.

It follows that selecting groups on the basis of their X-ray emission
will generate samples that, compared to a FoF selection from galaxy catalogues, are dominated by dynamically evolved systems. It will also tend to under-represent groups that have experienced strong feedback.

Of course, optical selection of groups introduces its own biases.
As has been well discussed in the literature \citep[e.g.][]{Gladders:2000aa, McNamara:2001aa, Barkhouse:2006aa}. FoF-selected groups may be dynamically young to the point where they are not even gravitationally bound, and such catalogues are vulnerable to contamination by line-of-sight groupings, as well as to groups created by fragmentation of larger structures by an FoF algorithm. X-ray emission is less vulnerable to such effects due to the squared dependence of X-ray emissivity on gas density, which causes X-ray emission to pick out the deepest parts of a gravitational potential \citep[e.g.][]{Gal:2006aa, Finoguenov:2009aa}.

In the case of our own study, it may be of interest to quote some basic statistics to give an idea of the relative power of the GAMA and XXL surveys for detecting galaxy groups.
Excluding the high background XXL fields, the total 
area covered by our survey is $\sim17$ deg$^2$. In this area we have a total of 230 GAMA groups with five or more galaxy members, although we have excluded some of these as being possibly associated with complex larger structures, and our cuts in redshift and group optical luminosity, together with the removal of groups with high X-ray upper limits, reduced the final sample analysed in detail to 142 groups. Our X-ray photometry provides detections (in the sense that our X-ray posterior distribution has a non-zero peak) for 131 of these 142 groups.

For comparison, the XXL catalogue of \citetalias{Adami:2018aa} contains 139 galaxy groups and clusters detected in X-rays within the same area of sky. Of these, only 94 groups lie below a redshift of $z=0.45$, the maximum redshift of the GAMA group catalogue. This includes high statistical quality sources (designated `C1 clusters' in the XXL project) together with lower statistical quality (C2 and C3) sources that have been confirmed to be clusters by optical spectroscopy.
We find optical GAMA group counterparts in our sample (within 2 Mpc and 5000 km s$^{-1}$) for 64 of these X-ray sources (including cases where two X-ray sources may match a single optical FoF group). These figures show how well-matched the GAMA and XXL surveys are for a study of this kind.

Of the 30 X-ray groups that have no matches to optical groups in our sample, 18 have redshifts above $z=0.3$. It is likely that many of these X-ray systems have a corresponding optical group too distant for the GAMA survey to detect at least five members. To investigate this, we checked for optical counterparts to each XXL source, this time including groups with multiplicities of at least two members. This resulted in all but five sources having an optical counterpart within 2 Mpc and 5000 km s$^{-1}$. These remaining five sources have redshifts $z>0.34$, so either they are too distant for GAMA to detect potential group members, or the GAMA FoF algorithm has incorporated them into larger structures that do not match within 2 Mpc.

\subsection{The role of large-scale structure}
One possibility that has not been explored in the present study is that some differences in X-ray emission may be related to differences in the large-scale structure surrounding groups. Filaments and other large-scale structures are known to change the properties of galaxies entering group and cluster environments \citep[e.g.][]{Porter:2008aa, Martinez:2016aa, Kleiner:2017aa, Kraljic:2018aa}. It could be that X-ray overluminous groups are able to accrete more gas from structures such as filaments, increasing the density of the IGM and thus enhancing their X-ray emission. Additionally, proximity to larger structures may also explain the properties of the EXdim systems. The present study removed several groups embedded in larger cluster systems and an analysis of the large-scale environment of the group sub-samples has not been completed. This hypothesis will be examined in a future publication.

\subsection{Potential biases}
\label{Section:biases}
Here we address some of the biases that may affect our results, and examine the sensitivity of our results to these factors.

We have already discussed the potential bias that could arise from the difference in L$_{opt}$ between our three sub-samples, which means that our results could be partially driven by L$_{opt}$. 
However, the X-ray overluminous groups in our sample have a {lower} L$_{opt}$ than the X-ray normal or underluminous groups. Previous studies have shown that the fraction of red galaxies, and the magnitude gap both tend to increase with richness, and hence with L$_{opt}$, so that a larger group should have a higher fraction of red galaxies, and a larger magnitude gap \citep[e.g.][]{Weinmann:2006aa, Davies:2019aa}. 
Our results contradict this expectation, since our overluminous groups tend to have the largest magnitude gaps, and highest passive fractions. Hence these results cannot arise from differences in L$_{opt}$ between X-ray over- and underluminous sub-samples, but must indicate genuine evolutionary effects.

Secondly, whilst the optical luminosities of our groups are well constrained, there are a number of effects that might affect the reliability of our X-ray measurements. Some of the X-ray flux estimates are subject to substantial statistical uncertainty. Hence a fraction of our points will have scattered from one sub-sample to the next. To check the impact of this, we artificially alter the X-ray flux measurement of each group in our sample, multiplying the existing value by the corresponding one sigma error multiplied by a random number drawn from a standard normal curve. In this way we produce 100 randomly perturbed datasets. We find that an average of 22 non-upper-limit groups change sub-sample as a result of these perturbations. 
This corresponds to less than $20\%$ of our sample changing sub-samples. This is unlikely to have a major impact on our results, although such statistical `blurring' will tend to reduce the significance of real trends seen in the data.

One advantage of the XMM EPIC instrument is that it has three independent CCD cameras, known as MOS1, MOS2, and PN Hence, we can derive three separate flux estimates for a given source and compare them. 
We used this to check that the flux estimates from all three cameras were in reasonable agreement and found nine groups in which results from one camera differed from the other two by $>3.5\sigma$ for no obvious reason. This level of disagreement is such that one would not expect even one example to arise by chance in our sample. Since we have identified no specific problem with these discrepant measurements we have  removed them from our analysis. However, to check the effects of discarding these discrepant readings, we repeated our analysis, omitting the discrepant count rate measurements for these nine groups and using the remaining cameras for our flux estimate. The revised X-ray fluxes causes five of these groups to change sub-sample (e.g. from X-ray underluminous to X-ray normal). However, this produces no qualitative difference to any of our principal results.

During this analysis we also identified three groups with significant masking from excised point sources or CCD chip gaps in the inner region of the source circle. These measurements had less than $50\%$ of the aperture removed by masking, and so were not discarded by our automatic requirement (see Sect.  \ref{Section:XrayPhotometry}) that only cameras with $>50\%$ of the source circle unmasked are used for flux estimation. However, there is still a danger that fluxes could be underestimated in these cases. We therefore repeated the analysis, discarding measurements that were manually flagged as having significant CCD gaps or masked central point sources. None of these three sources changed sub-samples with the revised X-ray fluxes, and the changes had no impact on our results. 

Additionally, point sources masked in our X-ray photometry followed the point source masks of \citet[\citetalias{Faccioli:2018aa}]{Faccioli:2018aa}. Whilst most of the sources in the masks are known to be point sources, it is possible that some of the masked X-ray emission could be from faint group emission. Exclusion of these sources could create an underestimate of the X-ray flux, as reported in \citet{Willis:2021aa}. However, as mentioned in Sect. \ref{Section:XrayPhotometry}, we find no correlation between the X-ray sub-sample with the number of masked point sources within 150 kpc of the group location, or with the masked area fraction. While it may be possible that some group flux may be masked out in a small number of cases, this does not affect our results in a systematic way.

\citet{Willis:2021aa} also noted that a larger distance from an XMM field centre can lower the probability of detecting a source. It could be possible that many X-ray underluminous groups have a high off-axis angle in the XMM field, causing the difference in R$_{XO}$. We compare the off-axis angles of all three sub-samples, and find that the distributions are not significantly different, with an A-D test having a value of p=0.19. This shows that our results are unlikely driven by differences in the XMM field location.

A third factor that may affect our results is the choice of the threshold lines in the L$^{XXL}_{300kpc}$ versus L$_{opt}$ plane that separate our three sub-samples in Fig. \ref{fig:Xraycuts_final}. To investigate this we tried adjusting both the slope and the normalisation of our two threshold lines.
We increased and decreased the power law slope by $0.5$, re-scaling the normalisation of the line to intersect the original fit at the median L$_{opt}$ value. We then used these adjusted power laws to redefine our three sub-samples, and compared the sub-sample sizes to our original sub-samples. The number of galaxies in any of the three sub-samples changed by no more than $4\%$ when increasing the power law exponent by 0.5, and $5\%$ when decreasing the exponent by 0.5.

The normalisation of the lines used to separate the three sub-samples -- two times above the scaling relation and three times below -- were relatively arbitrary. These factors were chosen to be round numbers, which resulted in similar sized overluminous and underluminous sub-samples. 
Again, we have checked that our results are robust to modest changes to these factors, though of course the statistical significance of our trends
is reduced if the factors are made too small (such that the subsets
are not clearly separated in R$_{XO}$) or too large (resulting in small sub-samples of over- and underluminous systems).

Finally, we repeated the analysis without excluding the potentially `complex' systems. In our original selection of groups for the study, we omitted 13 groups with multiple close X-ray sources, arguing that these could be merging or interacting in larger scale structures. Again we find that including these systems results in no significant change to our main results. 

\section{Summary}
\label{Section:Summary}
In this study we have combined data from overlapping high quality surveys in the X-ray and optical bands to generate a sample of galaxy groups with good
data in both bands, which has not been available previously. Combining
medium-deep X-ray data taken over a large contiguous area for the XXL cluster survey with spectroscopically selected groups from the GAMA survey has allowed
the assembly of a large sample of optically selected groups with matched 
X-ray flux values or upper limits. 

Using this sample we have compared groups with strong and weak X-ray emission
to investigate the relationship between the presence of hot X-ray emitting
gas and other group properties. We separated our group sample into subsets
with low, medium, and high X-ray emission for their mass, based
on the parameter R$_{XO}=L_{X300kpc}/L_{opt}^{2.03}$, and found the following:

\begin{itemize}

  \item Groups that are X-ray overluminous (high R$_{XO}$) have a BGG offset from the optical luminosity weighted centre that is 1.8 times lower, and an $r$-band magnitude gap between the first and second ranked galaxies on average 0.22 mag higher, than groups that are X-ray underluminous. 
  \item X-ray overluminous groups also contain a higher fraction of passive galaxies compared with X-ray underluminous systems, with a star forming fraction 1.7 times lower in the overluminous sub-sample.
  \item Trends in group structural parameters between the three sub-samples are less marked, possibly due to the large statistical errors associated with limited numbers of member galaxies. However, restricting the sample to groups with at least eight members indicates that X-ray overluminous groups have a higher dynamical mass-to-light ratio compared with X-ray underluminous groups.
  \item These results are consistent with the hypothesis that X-ray luminosity is an indicator of evolutionary maturity and that most systems with low R$_{XO}$ are not yet fully virialised.
  \item However, approximately 20\% of our X-ray underluminous sub-sample seems to consist of groups that are dynamically evolved. These groups do not appear to have had their gas exhausted by exceptionally high efficiency star formation, so we conclude that the most likely explanation for their low X-ray luminosity is a high entropy IGM, probably the result of vigorous feedback.
\end{itemize}

Future work to solidify our result that the group evolutionary stage is a key driver of the properties of most groups will likely require further observations. While some of the marginal trends in group structural parameters could be clarified with more spectroscopic characterisation of our existing group sample, a greater total number of groups would be beneficial. X-ray coverage of another GAMA field is being obtained as part of the eROSITA Final Equatorial Depth Survey (eFEDS) programme with the \textit{eROSITA} satellite \citep{Predehl:2021aa, Brunner:2021aa}. However, these observations are substantially less deep than the XXL data and have poorer spatial resolution. Hence, whilst they may provide useful constraints on the mean trends of X-ray luminosity with optical group properties (e.g. by stacking the X-ray data), they will not have sufficient depth to investigate the {scatter} in properties, as we have done in the present study. Thus, in the short term a combination of \textit{eROSITA} coverage and pointed X-ray follow-up of the well-characterised GAMA groups is the best way to increase the sample. In the future, large field spectroscopic surveys such as the 4most Hemisphere Survey with the 4-metre Multi-Object Spectroscopic Telescope \citep[4MOST,][]{deJong:2019aa} will hugely increase the number of groups with GAMA-like characterisation, but the lack of XXL-depth X-ray coverage over the field will be the limiting factor for at least the next 10 years. 

The most immediate route for further progress is a follow-up of our sample of dynamically evolved X-ray underluminous galaxy groups. These systems may already be the ‘missing link’ in the feedback cycle of groups. 

\begin{acknowledgements}
The authors thank the anonymous referee for their suggestions which improved the paper. We also wish to thank L. Chiappetti and E. Pompei for their helpful comments and feedback. 
JPC and SLM acknowledge support from the Science and Technology Facilities Council through grant number ST/N000633/1. JPC acknowledges support from the University of Birmingham, as well as support from Comit\'e Mixto ESO-Gobierno de Chile. MP thanks the Centre National d'Etudes Spatiales for long-term support.

GAMA is a joint European-Australasian project based around a spectroscopic campaign using the Anglo-Australian Telescope. The GAMA input catalogue is based on data taken from the Sloan Digital Sky Survey and the UKIRT Infrared Deep Sky Survey. Complementary imaging of the GAMA regions is being obtained by a number of independent survey programmes including GALEX MIS, VST KiDS, VISTA VIKING, WISE, Herschel-ATLAS, GMRT and ASKAP providing UV to radio coverage. GAMA is funded by the STFC (UK), the ARC (Australia), the AAO, and the participating institutions. The GAMA website is \url{http://www.gama-survey.org/}.

Based on observations obtained with XMM-Newton, an ESA science mission with instruments and contributions directly funded by ESA Member States and NASA. 
XXL is an international project based around an XMM Very Large Programme surveying two 25 deg2 extragalactic fields at a depth of $\sim 6 \times 10^{-15} $ erg cm$^{-2}$ s$^{-1}$ in the [$0.5-2$]~keV band for point-like sources. The XXL website is \url{http://irfu.cea.fr/xxl}. Multiband information and spectroscopic follow-up of the X-ray sources are obtained through a number of survey programmes, summarised at \url{http://xxlmultiwave.pbworks.com/}.
\end{acknowledgements}

\bibpunct{(}{)}{;}{a}{}{,} 
\bibliographystyle{aa} 
\bibliography{JCrossett_XXL_GAMA_bib.bib}

\begin{appendix}
\onecolumn

\section{List of groups used in this study.}

\begin{table}[h]
\centering
\caption{All groups used in this study. The full table is available online.} 
\begin{tabular}{c c c c c c c}

\hline
Group ID & RA & Dec & z & L$_{opt}$ & F$^{XXL}_{300kpc}$ & notes \\[0.2cm]
& (deg) & (deg) & & ($10^{11}$ L$_{\odot}/h^{2}$) & ($10^{-16} $erg s$^{-1}$cm$^{-2}$) & \\[0.2cm]
\hline
400014 & 30.9770 & -4.2325 & 0.135 & 2.39 & $257.0^{46.9}_{46.9}$& - \\ [0.25cm]
400019 & 37.8915 & -5.2619 & 0.142 & 2.11 & $115.0^{40.8}_{38.3}$ & - \\ [0.25cm]
400026 & 36.5726 & -5.0787 & 0.053 & 2.43 & $1410.0^{105.0}_{105.0}$ & removed due to substructure \\ [0.25cm]
400028 & 31.6183 & -4.6862 & 0.135 & 1.47 & $21.6^{66.9}_{15.2}$ & - \\ [0.25cm]
400029 & 30.8175 & -5.4053 & 0.210 & 2.29 & $93.1^{57.9}_{47.9}$ & - \\ [0.25cm]
400032 & 36.6178 & -4.0144 & 0.208 & 1.75 & $40.9^{45.9}_{25.9}$ & - \\ [0.25cm]
400039 & 36.8133 & -5.0756 & 0.143 & 1.81 & $0.0^{27.1}_{0.0}$ & - \\ [0.25cm]
\hline
\end{tabular}
\tablefoot{
Columns: 1) The Group ID as listed in the GAMA survey 2-3) The right ascension and declination of each group as determined by the iterative centre as per \citet{Robotham:2011aa}, unless otherwise noted 4) The median redshift of the group 5) The total redshift corrected group optical luminosity as per \citet{Robotham:2011aa} 6) The 300 kpc XXL X-ray flux values, with positive and negative $1\sigma$ uncertainties shown (described in Sect. \ref{Section:XrayPhotometry}) 7) Notes describing any alterations made to the data, and if the group was removed from our sample for any reason (see Table \ref{table:Xray_samples} for details). 
}
\end{table}

\label{lastpage}
\end{appendix}
\end{document}